\documentclass[12pt,preprint]{aastex}
\shorttitle{}
\shortauthors{}
\begin{document}
\baselineskip 19.pt
\title{Dynamical Model for the Zodiacal Cloud and Sporadic Meteors}
\author{David Nesvorn\'y$^1$, Diego Janches$^2$, David Vokrouhlick\'y$^{1,3}$, 
Petr Pokorn\'y$^{1,3}$, \\William F. Bottke$^1$, and Peter Jenniskens$^4$, } 
\affil{(1) Department of Space Studies, Southwest Research Institute,  
1050 Walnut St., Suite~300, Boulder, Colorado 80302, USA}
\affil{(2) Space Weather Laboratory, Code 674, GSFC/NASA, \\Greenbelt, 
MD 20771, USA}
\affil{(3) Institute of Astronomy, Charles University, \\
V Hole\v{s}ovi\v{c}k\'ach 2, CZ-18000, Prague 8, Czech Republic}
\affil{(4) Carl Sagan Center, SETI Institute, \\515 N. Whisman Road, 
Mountain View, CA 94043, USA}

\begin{abstract}
The solar system is dusty, and would become dustier over time as asteroids collide and comets disintegrate, 
except that small debris particles in interplanetary space do not last long. They can be ejected from the 
solar system by Jupiter, thermally destroyed near the Sun, or physically disrupted by collisions. Also, 
some are swept by the Earth (and other planets), producing meteors. Here we develop a dynamical model for 
the solar system meteoroids and use it to explain meteor radar observations. We find that the Jupiter Family 
Comets (JFCs) are the main source of the prominent concentrations of meteors arriving to the Earth from the 
helion and antihelion directions. To match the radiant and orbit distributions, as measured by the Canadian 
Meteor Orbit Radar (CMOR) and Advanced Meteor Orbit Radar (AMOR), our model implies that comets, and JFCs in 
particular, must frequently disintegrate when reaching orbits with low perihelion distance. Also, the 
collisional lifetimes of millimeter particles may be longer ($\gtrsim10^5$ yr at 1~AU) than postulated in the 
standard collisional models ($\sim10^4$ yr at 1 AU), perhaps because these chondrule-sized meteoroids are 
stronger than thought before. Using observations of the Infrared Astronomical Satellite (IRAS) to calibrate the 
model, we find that the total cross section and mass of small meteoroids in the inner solar system are 
$(1.7$-$3.5)\times10^{11}$ km$^2$ and $\sim4\times10^{19}$ g, respectively, in a good agreement with previous 
studies. The mass input required to keep the Zodiacal Cloud (ZC) in a steady state is estimated to be 
$\sim10^4$-$10^5$ kg s$^{-1}$. The input is up to $\sim$10 times larger than found previously, mainly because 
particles released closer to the Sun have shorter collisional lifetimes, and need to be supplied at a faster 
rate. The total mass accreted by the Earth in particles between diameters $D=5$ $\mu$m and 1 cm is found 
to be $\sim$15,000 tons yr$^{-1}$ (factor of 2 uncertainty), which is a large share of the accretion flux 
measured by the Long Term Duration Facility (LDEF). Majority of JFC particles plunge into the upper atmosphere 
at $<$15 km s$^{-1}$ speeds, should survive the atmospheric entry, and can produce micrometeorite falls. 
This could explain the compositional similarity of samples collected in the Antarctic ice and stratosphere, 
and those brought from comet Wild 2 by the Stardust spacecraft. Meteor radars such as CMOR and AMOR see only 
a fraction of the accretion flux ($\sim$1-10\% and $\sim$10-50\%, respectively), because small particles impacting 
at low speeds produce ionization levels that are below these radars' detection capabilities.    
\end{abstract}
\section{Introduction}
The Zodiacal Cloud (ZC) is a circumsolar disk of small debris particles produced by asteroid 
collisions and comets. Nesvorn\'y et al. (2010, hereafter N10) developed a dynamical model for 
particle populations released by asteroids and comets, and used the model to determine the 
relative contribution of asteroid and cometary material to the ZC. They found that 
the mid-infrared (MIR) emission from particles produced in the asteroid belt is mostly confined 
to within latitudes $b\lesssim30^\circ$ of the ecliptic. Conversely, the ZC has a 
broad latitudinal distribution so that strong thermal emission is observed even in the direction 
to the ecliptic poles (e.g., Hauser et al. 1984, Kelsall et al. 1998). This shows that 
asteroidal particles can represent only a small fraction of the ZC.

Based on a comparison of the model with observations of the Infrared Astronomical 
Satellite (IRAS), N10 proposed that $\gtrsim$90\% of the ZC's emission at MIR wavelengths comes 
from dust grains released by Jupiter-Family Comets (JFCs), and $\lesssim$10\% comes from the Oort 
Cloud Comets (OCCs), Halley-Type Comets (HTCs) and/or asteroid collisions. In addition, it was
found that the mass input required to keep the ZC in a steady state largely exceeds the mass loss
in JFCs due to their normal activity (e.g., Reach et al. 2007). To resolve this problem, 
N10 suggested that the dominant mass fraction is supplied to the ZC by spontaneous 
{\it disruptions/splittings} of JFCs (e.g., Fern\'andez 2005, Di Sisto et al. 2009). 

N10's model implies that the orbits of small meteoroids (diameters $D\lesssim100$ $\mu$m) released by 
JFCs become significantly circularized by Poynting-Robertson (P-R) drag before they can reach $a \sim 
1$ AU, and contribute to the Earth impact record. These particles, at the time of their accretion by 
the Earth, should thus have relatively low impact speeds ($v<20$ km s$^{-1}$), low eccentricities, 
and $a\sim1$ AU. The large JFC particles ($D\gtrsim1$ mm), on the other hand, should have a broader distribution 
of impact speeds, large eccentricities, and $a\sim2$-4 AU, mainly because they have presumably short collisional 
lifetimes (Gr\"un et al. 1985; hereafter G85), 
and disrupt before they can significantly evolve by P-R drag. As we discuss below, these 
results appear to be at odds with the observations of sporadic meteors.

Meteors are produced by small interplanetary particles also known as the {\it meteoroids}. Based on meteor 
data, the meteoroids can be divided into two groups: sporadic meteoroids and meteoroid streams. The 
meteoroid streams are prominent concentrations of particles with similar orbits (Whipple 1951). 
They are thought to be produced by particles released by active and recently ($<$ few thousand years ago) disrupted 
comets (Jenniskens 2008). Sporadic meteoroids are those particles that have evolved significantly 
from their parent body so that they are no longer easily linked to that parent, or to other meteoroids 
from the same parent. Notably, the time-integrated flux of meteors at Earth is dominated by about a factor 
of $\sim$10 by sporadics (Jones \& Brown 1993).

The radiant distribution of sporadic meteors shows several concentrations on the sky known as the 
helion/antihelion, north/south apex, and north/south toroidal sources (e.g., Campbell-Brown 2008, and 
the references therein). The prominent helion/antihelion source is the concentration of meteors
near the helion and antihelion directions. These meteors are believed to originate from the same 
population of meteoroids. The two groups differ in impact direction because some particles will 
impact before their perihelion passage, thus producing meteors with the antihelion radiants, while others will 
impact after their perihelion passage, producing meteors with the helion radiants. 

The helion/antihelion 
meteoroids have a measured impact speed distribution that peaks at $v\simeq20$-30 km s$^{-1}$, $a\sim1$ AU
with a tail to 3 AU and beyond, $e>0.3$, and low inclinations (Fig. \ref{amor0}).  
Wiegert et al. (2009; hereafter W09) developed a dynamical model to explain these observations. They 
found that particles released by JFCs, mainly by 2P/Encke, provide the best match to the observed properties 
of the helion/antihelion source (see also Jones et al. 2001).\footnote{The north/south apex meteors are 
most likely produced by meteoroids released from retrograde HTCs and/or OCCs (Jones et al. 2001, W09, 
Nesvorn\'y et al. 2011). The origin of the toroidal source is unknown.}

Comet 2P/Encke has an orbit that is quite unique among JFCs ($a=2.2$ AU, $e=0.85$, $i=11.8^\circ$),
because its aphelion distance lies well within Jupiter's orbit ($Q=a(1+e)=4.1$ AU). The orbit is 
relatively stable as it is not affected by close encounters with 
Jupiter.\footnote{The dynamical origin of 2P/Encke has yet to be explained, but probably requires 
non-gravitational forces produced by jets of material escaping from the comet's surface, and gravitational 
perturbations from the terrestrial planets (Valsecchi 1999).} 
In addition, the comet has a very low perihelion distance ($q=a(1-e)=0.34$ AU) and is expected to fall into the 
Sun in $10^5$-$10^6$ years (Levison \& Duncan 1994), if it physically survived that long. Finally,
2P/Encke is the source of several meteor streams known as Taurids (Whipple 1939), suggesting,
as argued in W09, that it can also be an important source of sporadic meteoroids at 1 AU.

The difference between W09 and N10 lies, in part, in different assumptions on the initial distribution of 
meteoroids. In W09, the meteoroids launched from 2P/Encke initially had a low perihelion distance so that 
even after having evolved by P-R drag to $a\sim1$ AU, they still retained a relatively large eccentricity. 
In N10, on the other hand, most meteoroids were released with $q\sim2.5$ AU, and greater effects of P-R drag 
were thus required for particles to reach 1 AU. In addition, N10 did not properly include the detection
efficiency of meteor radars in their model. This is a central issue, because most meteor radars are only
capable of detecting the relatively large and/or fast meteoroids, and may thus produce measurements of the 
Earth accretion flux that are heavily biased by their detection capabilities.
 
The agreement between the W09 model and observations of helion/antihelion meteors is not perfect.
For example, the W09 model produced tightly clustered distributions of $v$ and $e$ about $v=30$ 
km s$^{-1}$ and $e=0.85$, and lacked orbits with $a\gtrsim2.5$ AU (Fig. 2 in W09). The observed speeds 
and eccentricities have larger spreads, perhaps indicating that the helion/antihelion meteors are produced 
not by one, but many parent comets with a broad distribution of orbits, including those with $a\gtrsim2.5$ AU.
The inclination distribution produced in W09 does not have the resolution needed for a careful comparison 
with data, but it also seems to be narrower than the observed distribution. 

While 2P/Encke, or orbitally similar comets, can be an important source for helion/anti\-helion 
meteors, comet 2P/Encke itself cannot be a single dominant source of the ZC. This is because studies 
of the ZC indicate that $\sim$1000 kg s$^{-1}$ of material need to be injected into interplanetary space 
to keep the ZC in a steady state (e.g., Leinert et al. 1983, N10). Also, according to N10, the present mass 
of the inner ZC at $<$5 AU is $\sim$1-$2\times10^{19}$ g, which is roughly equivalent to that of a 25 km
diameter body. For comparison, the mass loss in comet 2P/Encke is only $\sim$26 kg s$^{-1}$, based on 
observations of its dust trail (Reach et al. 2007), and diameter of the nucleus is $\simeq$ 4.8 km 
(Fernandez et al. 2000, Boehnhardt et al. 2008).

According to N10, the dominant mass fraction is supplied to the ZC by spontaneous
disruptions/splittings of JFCs. Since meteoroids in the ZC are also expected to produce meteors, the meteor 
observations discussed above can place important constraints on the ZC's origin. To take advantage of 
these constraints, and motivated by the results discussed above, we modify N10's model to include a 
$q$-dependent meteoroid production rate,\footnote{Direct evidence for JFC disruptions at small 
heliocentric distances comes from the comparison of dynamical 
models of JFCs, which follow their transport from the trans-Neptunian region to the inner solar system, 
with observations (e.g., Levison \& Duncan 1997, Di Sisto et al. 2009). If the radial density of JFCs expected 
from the dynamical model, assuming no disruptions, is normalized so that it matches the observed (complete) 
sample of active comets with $q\simeq 1.5$ AU, it becomes apparent that the model density profile drops far 
too slowly for $q<1.5$ AU to match observations. This means that the comets with small $q$ values must 
disappear, due to physical effects, more quickly than those with large $q$ values. To match the observed 
profile, Di Sisto et al. (2009) suggested that the disruption probability of JFCs scales with $q$ as 
$q^{-\zeta}$, where $\zeta \simeq 0.5$-1.} and account for the detection efficiency of meteor surveys. 
We also improve N10's model to consider a continuous Size Frequency Distribution (SFD) of particles, 
and more precisely parametrize their collisional disruption in space. We show that, with these modifications
of the N10 model, the results match available constraints. We describe the new model in Section 2. The 
results are reported in Section 3. We estimate the ZC's cross section and mass, meteoroid production rate 
required to keep the ZC in a steady state, and the terrestrial accretion rate of interplanetary dust.  
\section{Model}
Our model includes the following parts: (1) particles of different sizes are released from JFCs 
(Section 2.1), (2) their orbits evolve under the influence of gravitational and radiation forces 
(Section 2.2), (3) some particles are thermally or collisionally destroyed (Section 2.3), (4) 
while in space, particles emit thermal radiation (Section 2.4), which (5) is detected by a 
telescope observing at MIR wavelengths (Section 2.5), and (6) a small fraction of the initial 
particle population is accreted by the Earth, producing meteors (Section 2.6). We describe 
components (1)-(6) below. 

Procedures described in Sections 2.5 and 2.6 are mainly required because the raw particle distributions 
obtained from our numerical integrations of orbits in Section 2.2 do not have a sufficient resolution. 
We use analytical methods to enhance the resolution in a way that is suitable for (5) and (6). 
\subsection{Initial Orbits}
We only consider JFCs in this paper, because previous works showed that they are the main source of 
the ZC particles and helion/antihelion meteors (e.g., Jones et al. 2001, W09, N10). The asteroid 
meteoroids have low impact speeds and are not detected by meteor radars. The meteoroids released from
long-period comets contribute to apex meteors and are not modeled here (see Nesvorn\'y et al. 
(2011) for a discussion of apex meteors).

The orbital distribution of JFCs was taken from Levison \& Duncan (1997; hereafter LD97), who 
followed the orbital evolution of bodies originating in the Kuiper belt as they are scattered by 
planets, and evolve in small fractions into the inner solar system. For each critical perihelion distance, 
$q^*$, we selected bodies from LD97's simulations when they reached 
$q<q^*$ for the first time. Particles were released from these source orbits.\footnote{Note that 
this method does not properly account for the possibility that JFCs lose mass gradually by 
recurrent splitting events. For example, Di Sisto et al. (2009) assumed that the splitting events
occur with certain frequency, considered to be a free parameter, and that a fixed mass fraction, 
also a free parameter, is lost in each event. Such a model may be physically more appropriate, but 
has more free parameters.}  We used 10 values of $q^*$ equally spaced between 0.25 AU and 2.5 AU, 
particles with $D=10$, 30, 100, 300, 1000, 3000 $\mu$m, which should cover the interesting range 
of sizes, and particle density $\rho=2$ g cm$^{-3}$. Our tests show that this size resolution is 
adequate because the orbital dynamics of, say, a $D=150$-$\mu$m particle is similar to that of
a $D=100$-$\mu$m particle. The results for a continuous range of sizes were obtained by 
interpolation. Varying particle density has only a small effect on their orbital dynamics.
Ten thousand particles were released for each $q^*$ and $D$ for the total of 0.6 million of the initial orbits.

Upon their release from the parent object, particles will feel the effects of radiation pressure. 
These effects can be best described by replacing the mass of the Sun, $m_\odot$, by $m_\odot(1-\beta)$, 
with $\beta$ given by
\begin{equation} \beta = 5.7\times10^{-5}\ {Q_{\rm pr} \over \rho s}\ , \label{beta} \end{equation} 
where the particle's radius $s=D/2$ and $\rho$ are in cgs units. Pressure coefficient $Q_{\rm pr}$ 
can be determined using the Mie theory (Burns et al. 1979). We set $Q_{\rm pr} = 1$, which corresponds 
to the geometrical optics limit, where $s$ is much larger than the incident-light wavelength. Note 
that all particles considered here are large enough to stay on bound heliocentric orbits after their
release (see, e.g., Nesvorn\'y et al. 2011).

In Section 3, we combine the results obtained with different $q^*$ and $D$ to mimic the continuous 
distributions ${\rm d}N(q)$ and ${\rm d}N(D)$. This is different from N10 where the source particle
distributions were parametrized by the `fading time', particle production rate was assumed to be 
$q$-independent, and ${\rm d}N(D)$ was approximated by single size. Here we consider
${\rm d}N(q)$ and ${\rm d}N(D)$ that can be approximated by simple power laws. For ${\rm d}N(q)$,
we thus have ${\rm d}N(q) \propto q^\gamma {\rm d}q$, where $\gamma$ is a free parameter. For 
${\rm d}N(D)$, we have ${\rm d}N(D)=N_0 D^{-\alpha} {\rm d}D$, where $N_0$ is a normalization constant
and $\alpha$ is the usual slope index (at source). Alternatively, we use the two-slope SFD
with ${\rm d}N(D) \propto D^{-\alpha_1} {\rm d}D$ for $D<D^*$ and ${\rm d}N(D) \propto D^{-\alpha_2} 
{\rm d}D$ for $D>D^*$, where $\alpha_1$, $\alpha_2$ and $D^*$ are free parameters. 

Parameter $\gamma$ can be inferred from the number of JFCs found at each $q$, and their disruption
probability as a function of $q$. If the former can be approximated by $q^{\xi}$, where 
$\xi \sim 0.5$ (LD97, Di Sisto et al. 2009), and the latter is proportional to $q^{-\zeta}$, where $\zeta \simeq 
0.5$-1 (Di Sisto et al. 2009), it would be expected that $\gamma=\xi-\zeta \sim -0.5$-0. We use 
$\gamma=0$ as our starting value, and test the sensitivity of results for $\gamma<0$ and $\gamma>0$.
As for ${\rm d}N(D)$, we set $\alpha \sim 4$ for the whole size range, as motivated by 
meteor radar observations (e.g., Galligan \& Baggaley 2004), or $D^* \sim 100$ $\mu$m, $\alpha_1<3$ 
and $\alpha_2>4$, as motivated by space impact experiments (e.g., Love \& Brownlee 1993). Given 
the various uncertainties of these measurements (see, e.g., Mathews et al. 2001), we also test $D^* <
100$ $\mu$m and $D^* > 100$ $\mu$m.    
\subsection{Orbit Integration} 
The particle orbits were numerically integrated with the {\tt swift\_rmvs3} code (Levison \&
Duncan 1994), which is an efficient implementation of the Wisdom-Holman map (Wisdom \& Holman 1991) 
and which, in addition, can deal with close encounters between particles and planets. The radiation 
pressure and Poynting-Robertson (P-R) drag forces were inserted into the Keplerian and kick parts of 
the integrator, respectively. The change to the Keplerian part was trivially done by substituting  
$m_\odot$ by $m_\odot(1-\beta)$. We assumed that the solar-wind drag force has the same functional 
form as the P-R term and contributes by 30\% to the total drag intensity.

The code tracks the orbital evolution of a particle that revolves around the Sun and is 
subject to the gravitational perturbations of seven planets (Venus to Neptune; the mass 
of Mercury was added to the Sun) until the particle impacts a planet, is ejected from the Solar 
System, evolves to within 0.05~AU from the Sun, or the integration time reaches 5 Myr. 
We removed particles that evolved to $R<0.05$~AU, because the orbital period for $R<0.05$~AU was 
not properly resolved by our 1-day integration timestep.\footnote{We tested an integration
timestep of 0.3 day. The results were essentially identical to those obtained with the 1-day timestep.} 
The particle orbits were recorded at 1,000-yr time intervals to be used for further analysis. 
\subsection{Physical Effects}
The solar system meteoroids can be destroyed by collisions with other particles and by solar heating 
that can lead to sublimation and vaporisation of minerals. Here we explain how we parametrize these 
processes in our model. 

The JFC particles will rapidly loose their volatile ices. We do not model the volatile loss here.
The remaining grains will be primarily composed from amorphous silicates and will survive down to 
very small heliocentric distances. Following Moro-Mart\'{\i}n \& Malhotra (2002), Kessler-Silacci 
et al. (2007) and others, we adopt a simple criterion for the silicate grain destruction. We 
assume that they are thermally destroyed (sublimate, vaporise) when the grain temperature reaches 
$T \geq 1500$ K.

Using the optical constants of amorphous pyroxene of approximately cosmic composition (Henning \& 
Mutschke 1997), we find that a dark $D\gtrsim100$-$\mu$m grain at $R$ has the equilibrium temperature 
within 10 K of a black body, $T \simeq 280/\sqrt{R}$ K. According to our simple destruction 
criterion, $T \geq 1500$ K, the silicate grains should thus be removed when reaching $R\lesssim0.035$ 
AU. On the other hand, the smallest particles considered in this work ($D=10$ $\mu$m) have
$T=1500$ K at $R\simeq 0.05$ AU. Thus, we opted for a simple criterion where particles 
of all sizes were instantly destroyed, and were not considered for statistics, when they reached 
$R \leq 0.05$ AU. Note that, by design, this limit is the same as the one imposed by the integration 
timestep (Section 2.2).          

The collisional lifetime of meteoroids, $\tau_{\rm coll}$, was taken from G85. It was assumed to be a 
function of particle mass, $m$, and orbital parameters, mainly $a$ and $e$. For example, for a circular 
orbit at 1 AU, particles with $D=100$ $\mu$m and 1 mm have $\tau^*_{\rm coll} = 1.5\times10^5$ yr and 
$7.3\times10^3$ yr, respectively, where $\tau^*_{\rm coll}$ denotes the collisional lifetime from G85.
Also, $\tau^*_{\rm coll}$ increases with $a$. 
To cope with the uncertainty of the G85's model, we introduced a free parameter, $S$, so that 
$\tau_{\rm coll} =S\tau^*_{\rm coll}$. Values $S>1$ increase $\tau_{\rm coll}$ relative to $\tau^*_{\rm coll}$, 
as expected, for example, if particles were stronger than assumed in G85, or if the measured impact fluxes 
were lower (e.g., Dikarev et~al. 2005, Drolshagen et~al. 2008).  See Nesvorn\'y et al. (2011) for a fuller description.  

Collisional disruption of particles was taken into account during processing the output 
from the numerical integration described in Section 2.2. To account for the stochastic nature 
of breakups, we determined the breakup probability $p_{\rm coll} = 1 - \exp(-h/\tau_{\rm coll})$, 
where $h=1000\, {\rm yr}$ is the output interval, and $\tau_{\rm coll}$ was computed 
individually for each particle's orbit. The code then generated a random number $0\leq x\leq1$, 
and eliminated the particle if $x<p_{\rm coll}$. 
 
We caution that our procedure does not take into account the small debris fragments that are 
generated by disruptions of larger particles. Instead, all fragments are removed from the system.
This is an important approximation whose validity needs to be tested in the future. 

\subsection{Thermal Emission of Particles} 
Meteoroids were assumed to be isothermal, rapidly rotating spheres. The absorption was 
assumed to occur into an effective cross section $\pi s^2$, and emission out of $4 \pi 
s^2$. The infrared flux density (per wavelength interval ${\rm d} \lambda$) per unit surface 
area at distance $r$ from a thermally radiating particle with radius $s$ is 
\begin{equation} F_\lambda = \epsilon(\lambda,s) B(\lambda,T) {s^2 \over r^2} \ , 
\label{flux} \end{equation} 
where  
$\epsilon$ is the emissivity and $B(\lambda,T)$ is the energy flux  at 
$(\lambda,\lambda+{\rm d}\lambda)$ per surface area from a black body at temperature 
$T$: 
\begin{equation} B(\lambda,T) = {2 \pi h c^2 \over \lambda^5} \left[ {\rm 
e}^{hc/\lambda k T} - 1 \right]^{-1} \ . \label{bb} \end{equation} 
In this equation, $h=6.6262\times10^{-34}$ J s is the Planck constant, $c = 2.99792458 
\times 10^8$ m s$^{-1}$ is the speed of light, and $k = 1.3807\times10^{-23}$ J K$^{-
1}$ is the Boltzmann constant. 

Since our model does not include detailed emissivity properties of dust grains at different 
wavelengths, we set the emissivity at 25 $\mu$m to be 1 and fit for the emissivities at 12 and 
60~$\mu$m. We found that the relative emissivities at 12 and 60 $\mu$m that match the data 
best are 0.70-0.75 and 0.95-1, respectively. Such a variability of MIR emissivity values 
at different wavelengths is expected for small silicate particles with some carbon content. 
$T(R)$ was set to be $280/\sqrt{R}$ K, as expected for dark $D\gtrsim10$-$\mu$m particles. See 
Nesvorn\'y et al. (2006) for a more precise treatment of $\epsilon(\lambda,s)$ and $T(R)$ for 
dust grains composed of different materials. 
\subsection{MIR Observations} 
To compare our results with IRAS observations illustrated in Fig. \ref{mean},\footnote{We use
IRAS because it is the dataset we are best familiar with (see Nesvorn\'y et al. 2006, N10).
Other, more modern MIR surveys such as the Cosmic Background Explorer (COBE; e.g., Kelsall et al. 1998) 
have better precision 
and resolution, but their results do not differ in important ways from those obtained by IRAS.
The COBE measurements of the extended MIR emission were used to calibrate the IRAS fluxes as described
in  Nesvorn\'y et al. (2006).} we developed a 
code that models thermal emission from distributions of orbitally evolving particles and 
produces infrared fluxes that a space-borne telescope would detect depending on its location, 
pointing direction and wavelength. See Nesvorn\'y et al. (2006) for a detailed description of 
the code. 

In brief, we define the brightness integral along the line of sight of an infrared 
telescope (defined by fixed longitude $l$ and latitude $b$ of the pointing 
direction) as:
\begin{equation}
\int_{a,e,i} {\rm d}a {\rm d}e {\rm d}i
\int_{0}^{\infty} {\rm d}r\ r^2 
\int_D {\rm d}D\, F_\lambda(D,r) N(D;a,e,i) K(R,L,B) \ ,
\label{all}
\end{equation}
where $r$ is the distance from the telescope, $F_\lambda(D,r)$ is the infrared flux 
(evaluated at the effective wavelength of the telescope's system) per unit surface 
area at distance $r$ from a thermally radiating particle with diameter $D$. 
$K(R,L,B)$ defines the spatial density of particles in sun-centered coordinates as 
a function of $R$, ecliptic longitude, $L$, and latitude, $B$. $N(D,a,e,i)$ is the number of 
particles having effective diameter $D$ and orbits with $a$, $e$ and $i$. 

We evaluate the integral in Eq. (\ref{all}) by numerical renormalization (see Nesvorn\'y et al. 2006). 
$F_\lambda(D,r)$ is calculated as described in Section 2.4. $N(D,a,e,i)$ is obtained from our numerical simulations 
(Section 2.2). $K(R,L,B)$ uses analytic expressions for the spatial distribution of particles with 
fixed $a$, $e$ and $i$, and randomized orbital longitudes (Kessler 1981). 

We assume that the telescope is located at $(x_t=r_{\rm t} \cos \phi_{\rm t},y_t=r_{\rm t} 
\sin \phi_{\rm t},z_{\rm t}=0)$ in the Sun-centered reference frame with $r_{\rm t} = 1$ AU. Its viewing 
direction is defined by a unit vector with components $(x_{\rm v},y_{\rm v},z_{\rm v})$. In Eq. (\ref{all}), 
the pointing vector can be also conveniently defined by longitude $l$ and latitude $b$ of the 
pointing direction, where $x_{\rm v} = \cos b \cos l$, $y_{\rm v} = \cos b \sin l$, and $z_{\rm v} = 
\sin b$. We fix the solar elongation $l_\odot = 90^\circ$, so that $l=\phi_{\rm t}+90^\circ$, and 
calculate the thermal flux of various particle populations as a function of $b$ and wavelength. The 
model brightness profiles at 12, 25 and 60 $\mu$m are then compared with the mean IRAS profiles
shown in Fig.~\ref{mean}. 
\subsection{Model for Meteor Radar Observations}
We used the \"Opik theory (\"Opik 1951) to estimate the expected terrestrial accretion 
rate of JFC particles in our model. Wetherill (1967), and later Greenberg (1982), improved 
the theory by extending it more rigorously to the case of two eccentric orbits. Here we used 
a computer code that employs Greenberg's formalism (Bottke et al. 1994).\footnote{The 
\"Opik theory cannot properly account for the capture 
of particles in orbital resonances (e.g., Dermott et al. 1994, \v{S}idlichovsk\'y \& 
Nesvorn\'y 1994). Testing the effect of orbital resonances on particles released by JFCs is 
left for future work.}  

We modified the code to compute the radiants of the impacting particles. In doing so we 
properly accounted for all impact configurations and weighted the results by the probability 
with which each individual configuration occurs, including focusing. The radiants were 
expressed in the coordinate system, where longitude $l$ was measured from the Earth's apex in 
counter-clockwise direction along the Earth's orbit, and latitude $b$ was 
measured relative to the Earth's orbital plane. Note that our definition of longitude differs 
from the one more commonly used for radar meteors, where the longitude is measured from the 
helion direction. The radiants were calculated before the effects of gravitational focusing 
were applied. 

The meteor radars use different detection methods (i.e., trail vs. echo)\footnote{Note that 
the parametrization described here applies to the specular meteor radars, which detect the meteor 
trails. A similar parametrization can be developed, however, for the more sensitive High Power 
and Large Aperture (HPLA) Radars (Fentzke \& Janches 2008, Fentzke et al. 2009)
that detect meteor head echoes.} and have different
sensitivities. Their detection efficiency is mainly a function of the particle's mass and 
speed, but it also depends on a number of other parameters discussed, for example, in Janches et al. 
(2008). Following W09, we opt for a simple parametrization of the radar sensitivity function, where 
the detection is represented by the ionization function
\begin{equation}
I(m,v) = {m \over 10^{-4}\, {\rm g}} \left( {v \over 30\,{\rm km/s}} \right)^{3.5} \ .  
\label{cut}
\end{equation}
All meteors with $I(m,v)\geq I^*$ are assumed to be detected in our model, while all meteors 
with $I(m,v) < I^*$ are not detected. The ionization cutoff $I^*$ is taken to be different for different radars.
For example, $I^*\sim 1$ for the Canadian Meteor Orbit Radar (CMOR; Campbell-Brown 2008) 
and $I^*\sim 0.01$-0.001 for AMOR (Galligan 
\& Baggaley 2004, 2005). For reference, a JFC particle with $v = 30$~km s$^{-1}$ and $m = 
10^{-6}$~g, corresponding to $D\simeq100$ $\mu$m with $\rho=2$ g cm$^{-3}$, has $I(m,v) = 0.01$, i.e., a 
value intermediate between the two thresholds. These meteoroids would thus be detected by AMOR,
according to our definition, but not by CMOR. The particle size detection threshold is shown, as a 
function of $v$, in Fig. \ref{icut}.

To study the orbital properties of different meteor sources, these sources need to be isolated.
This is typically done by selecting meteors with specific radiants. To test how the radiant
cutoff affects the results, we select the helion meteors with $-90^\circ<l<-45^\circ$ and 
$-30^\circ<b<30^\circ$, and antihelion meteors with $45^\circ<l<90^\circ$ and $-30^\circ<b<30^\circ$.
Since our code computes the same impact speed and orbit distributions for the helion and 
antihelion sources, we combine the results from the two radiant cutoffs together. Note, therefore,
that our method cannot capture the suspected asymmetry between the helion and antihelion sources 
(see, e.g., W09, and the references therein).
\section{Results}
We performed hundreds of tests with the model described in Section 2. The main parameters of
these tests were: the (1) size distribution of JFC particles at the source, as defined by $D^*$, 
$\alpha_1$ and $\alpha_2$, (2) power index of the initial perihelion distribution, ${\rm d}N(q) 
\propto q^\gamma {\rm d}q$, and (3) collisional lifetime of particles, $\tau_{\rm coll}$. To 
compare our model with meteor observations, we specified the appropriate ionization threshold, 
and applied the usual $\chi^2$ statistics (see, e.g., Nesvorn\'y et al. 2006). To simplify the 
presentation of results, we first discuss selected cases that illustrate the trends with different 
parameters. These cases are generally representative for a wide range of parameter values, as 
explained in the following text. 
 
\subsection{AMOR}
We start by discussing the results relevant to AMOR, because AMOR is capable of detecting 
particles with $D\sim100$ $\mu$m (Fig. \ref{icut}), and can thus provide constraints on the 
particle sizes that are thought to be dominant in the ZC. Figure \ref{amor1} shows the distributions
of impact speeds and orbits of JFC meteoroids for $D^*=100$ $\mu$m, $\gamma = 0$, $S=1$, 
and several values of the ionization cutoff.
With $I^*=0$, corresponding to no cutoff on mass or impact speed, the impact speed 
distribution, ${\rm d}N(v)$, is strongly peaked toward the Earth's escape speed ($v_{\rm esc}=11.2$ 
km s$^{-1}$). When $I^*=0.003$ cutoff is applied, as roughly expected for the AMOR detections, 
${\rm d}N(v)$ has a maximum at $v\simeq 25$~km~s$^{-1}$. {\it This illustrates the crucial importance 
of the ionization cutoff for the interpretation of meteor radar observations.} 

Given the strong effect of the ionization cutoff it is difficult to imagine how the radar 
observations can be correctly `debiased', based solely on the measurements, corrections, and 
considerations of the Earth-impact probability of different orbits (e.g., Taylor \& McBride 
1997, Galligan \& Baggaley 2004, Campbell-Brown 2008), to obtain the real distribution of meteoroids at 
1~AU. As shown in Fig. \ref{amor1}, the real distribution can be very different from the observed one; although 
meteors at low speeds may dominate the real distribution, only a tiny fraction are detected.
This highlights the importance of dynamical modeling.

In a similar fashion, the observed eccentricity distribution, ${\rm d}N(e)$, is strongly biased toward 
large values by the ionization cutoff, while the underlying distribution has more small and moderate values
(Fig. \ref{amor1}d). This explains, at least in part, why N10 were unable to obtain meteor-like 
${\rm d}N(e)$, because they did not model the meteor detection in detail. The model semimajor 
axis and inclination distributions, ${\rm d}N(a)$ and ${\rm d}N(i)$, are also strongly affected by 
the ionization cutoff. With $I^*=0.003$, both distributions become significantly broader than those 
computed for $I^*=0$ (Figs. \ref{amor1}bc).  

\begin{figure}
\epsscale{0.9}
\plotone{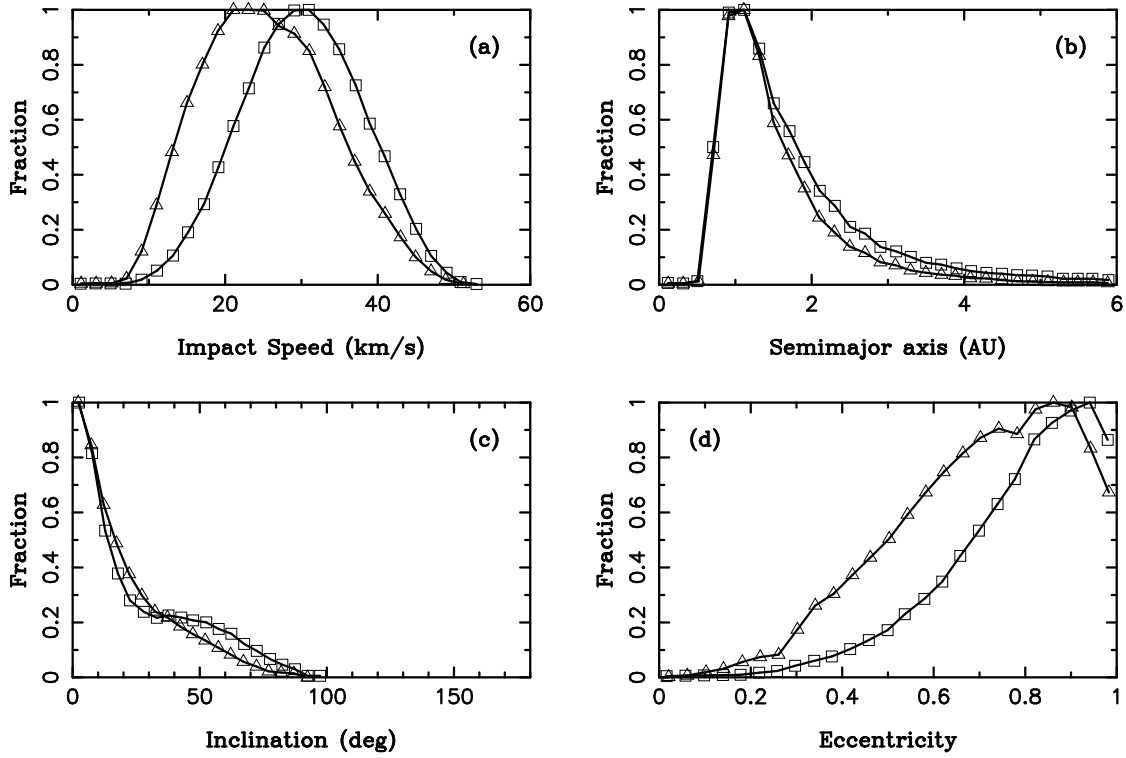}
\caption{The distributions of impact speeds and orbits of prograde antihelion meteors measured by
the Advanced Meteor Orbit Radar (AMOR). According to Galligan \& Baggaley (2005), the antihelion 
meteors were selected using a broad radiant cutoff ($20^\circ<l<120^\circ$, in our definition
of longitude --see Section 2.6--, and no condition on $b$). The squares label the raw distributions obtained 
by AMOR. The distributions labeled by triangles were corrected for the atmospheric interference
and Faraday rotation. The helion meteors, not shown here, have corrected distributions very similar 
to those plotted here. The impact speeds in panel (a) include effects of the gravitational focusing 
by Earth. The heliocentric orbital elements shown in (b), (c) and (d) do not include these effects.
All distributions were normalized to reach 1 at their maximum. Adapted from Fig. 8 in Galligan \& 
Baggaley (2005).}
\label{amor0}
\end{figure}

\begin{figure}
\epsscale{0.5}
\plotone{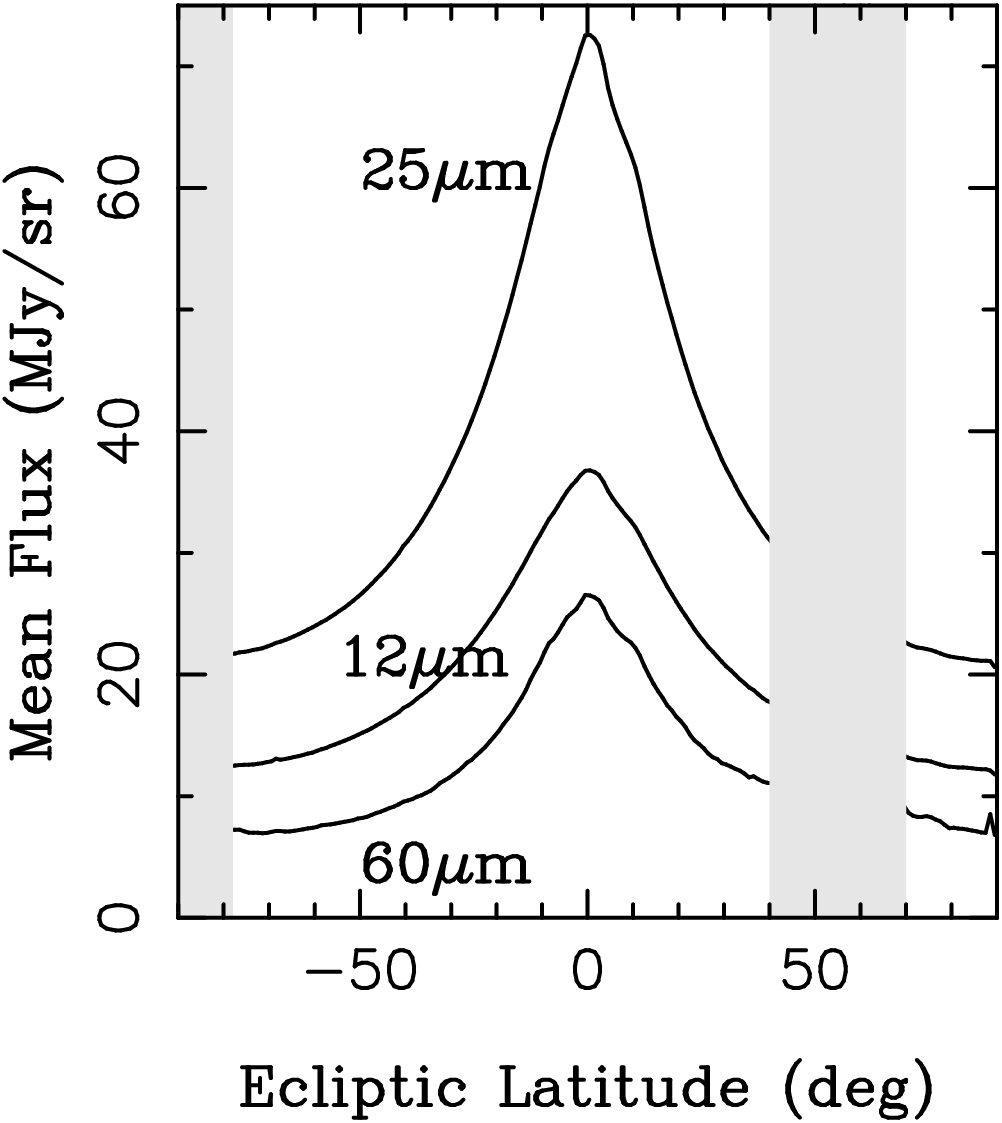}
\caption{Mean ZC profiles obtained by IRAS in 12, 25 and 60 $\mu$m wavelengths. To make these profiles, 
the selected IRAS scans obtained with the $\simeq90^\circ$ solar elongation were centered at the ecliptic, 
smoothed by a low-pass filter, and combined 
together (see N10 for details). The gray rectangles at $b<-78^\circ$ and $40^\circ<b<70^\circ$ block the latitude 
range where the mean fluxes were significantly affected by the galactic plane emission. We do not use 
the excluded range in this work. The uncertainties of the mean flux values are not shown here; they are 
too small to clearly appear in the plot. IRAS observations at 100 $\mu$m, not shown here, are less 
useful for probing the thermal radiation of dust particles in the inner solar system, because of 
the strong interference with the galactic and extra-galactic emission at these wavelenghts.}
\label{mean}
\end{figure}

\begin{figure}
\epsscale{0.5}
\plotone{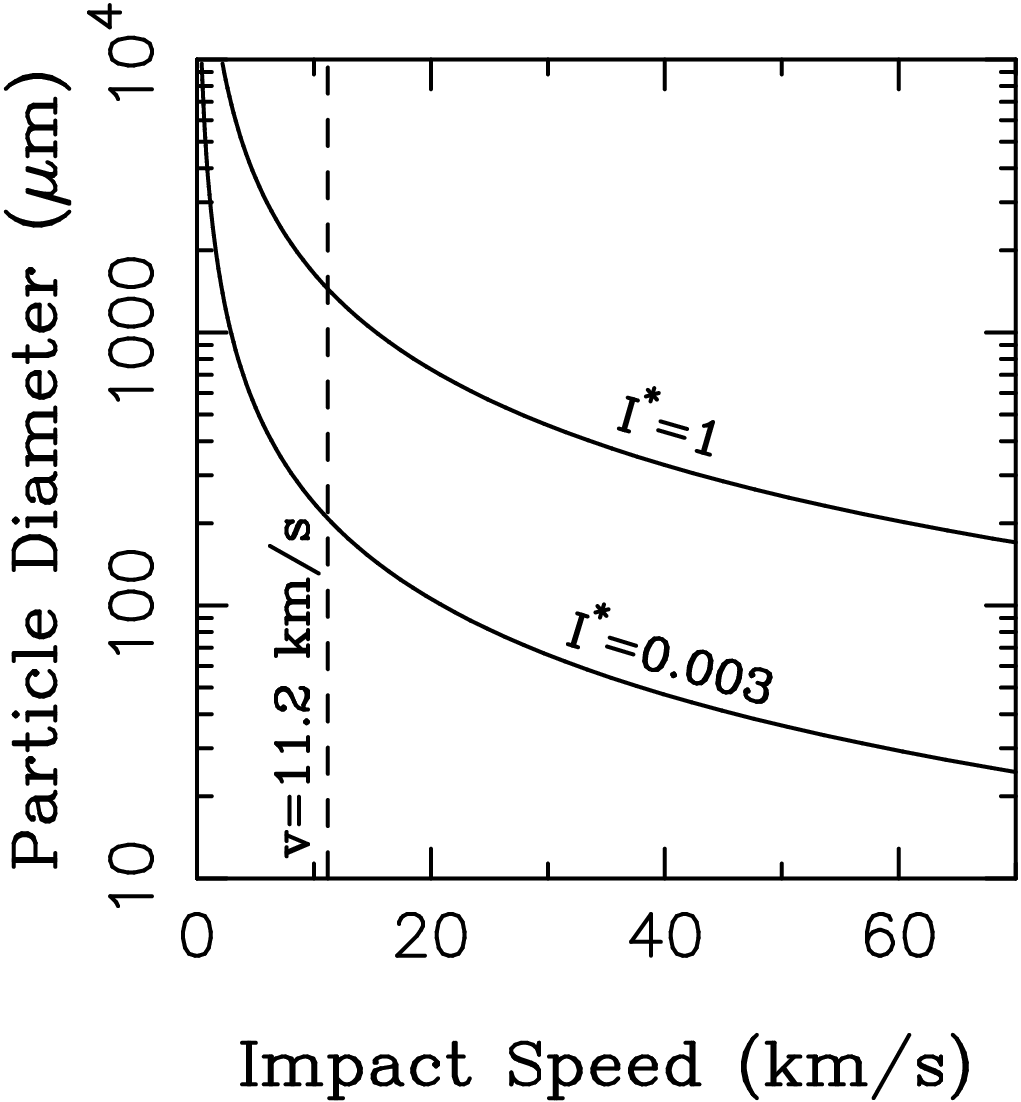}
\caption{The detection size threshold as a function of the meteor impact speed. All particles above the solid
lines are assumed to be detected. The thresholds are $I^*\simeq1$ for CMOR and $I^*\simeq0.003$ for 
AMOR. Meteors occur to the right from the dashed vertical line that denotes the Earth's escape speed 
($v_{\rm esc}=11.2$ km s$^{-1}$).}
\label{icut}
\end{figure}

\begin{figure}
\epsscale{0.9}
\plotone{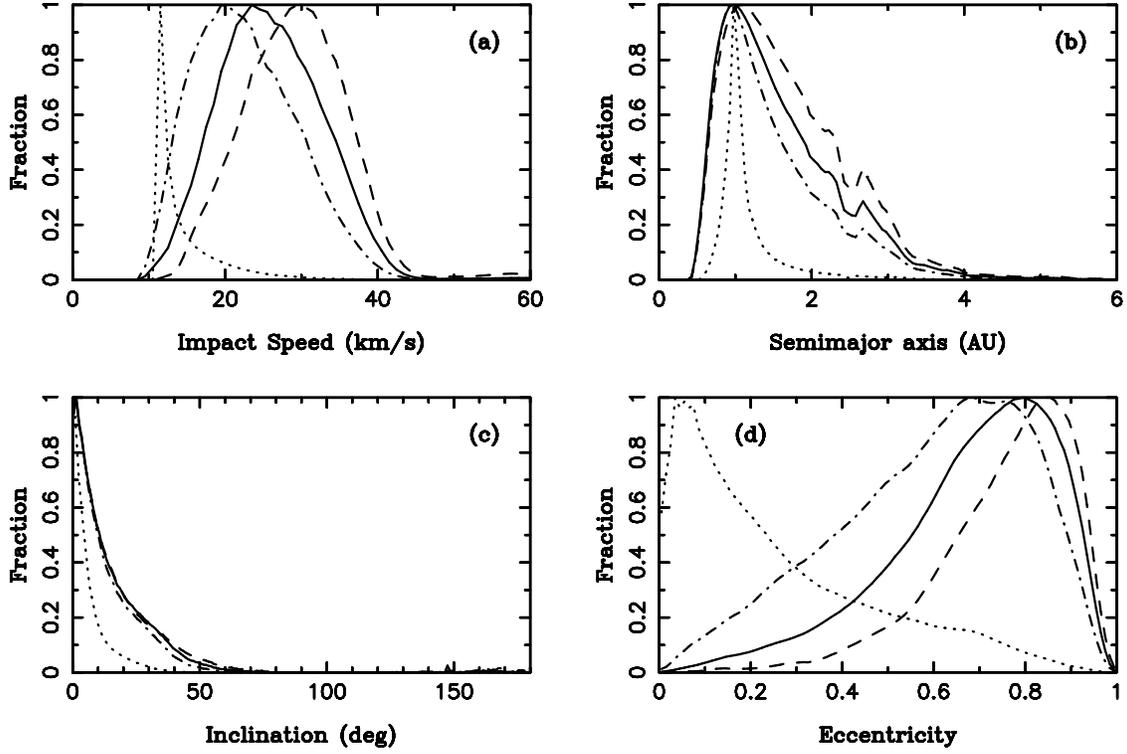}
\caption{Effect of the ionization cutoff. Different lines show the results for $I^*=0$ (dotted), 
$I^*=0.001$ (dot-dashed), $I^*=0.003$ (solid), and $I^*=0.01$ (dashed). As the ionization cutoff 
increases, the peak of ${\rm d}N(v)$ shifts to larger values. Here we used $D^*=100$~$\mu$m,
$\alpha_1=2$, $\alpha_2=5$, $\gamma=0$ and $S=1$. Most meteoroids accreted by Earth have $v < 15$ 
km s$^{-1}$ , while most meteoroids detected by AMOR have $v > 15$ km s$^{-1}$. No radiant cutoff was applied here.
The drop of ${\rm d}N(a)$ near $a=2.5$ AU corresponds to the gap in the distribution of orbits 
produced as particles drifting by P-R drag jump over the 3:1 mean motion resonance with 
Jupiter. All distributions were normalized to reach 1 at their maximum.}
\label{amor1}
\end{figure}

The distribution of impact speeds obtained in our model has a peak value of 
$v=20$ km s$^{-1}$ for $I^*=0.001$ and $v=30$ km s$^{-1}$ for $I^*=0.01$, in good agreement
with the AMOR measurements of helion/antihelion meteors that show a peak at $v=20$-25~km~s$^{-1}$.
The spread of model ${\rm d}N(v)$ (Fig. \ref{amor1}a), however, is slightly narrower 
than the one indicated by observations (Fig. \ref{amor0}a). We will discuss this difference later 
in this section, and show that it can be related to the initial SFD of particles produced by 
JFCs. 

\begin{figure}
\epsscale{0.9}
\plotone{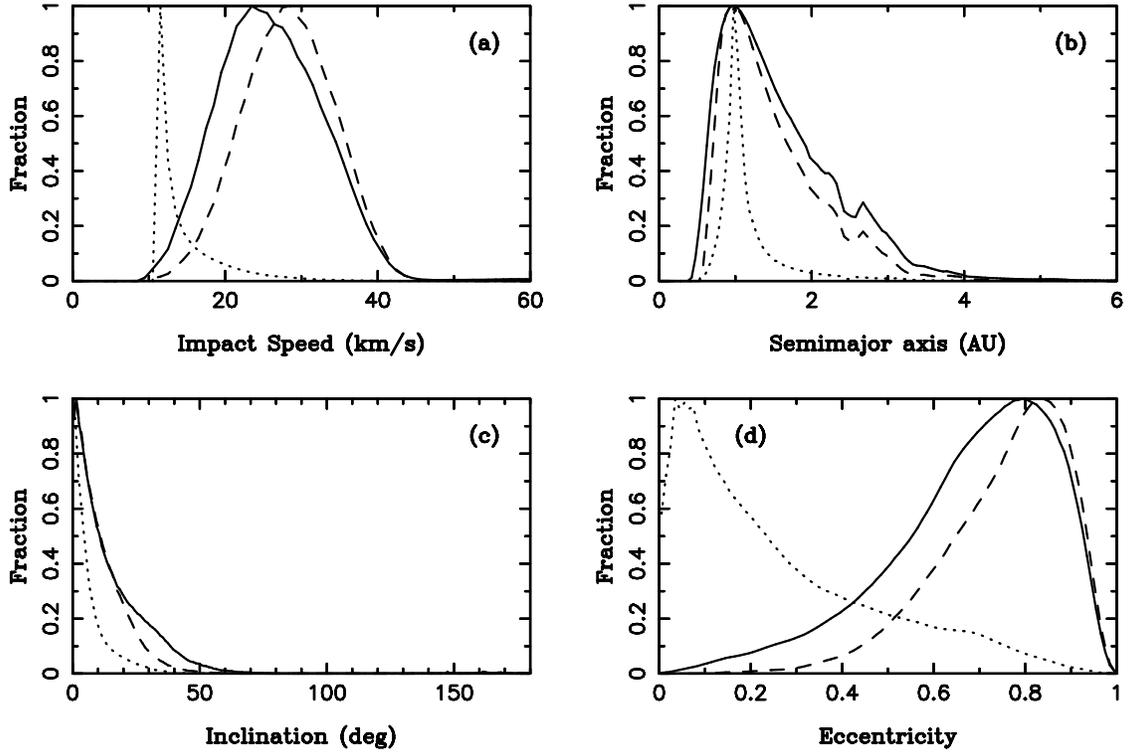}
\caption{Effect of the radiant cutoff. Different lines show the results for $I^*=0$ without a radiant cutoff 
(dotted), $I^*=0.003$ without a radiant cutoff (solid), and $I^*=0.003$ with radiant cutoff (dashed). As in 
Fig. \ref{amor1}, we used $D^*=100$ $\mu$m, $\alpha_1=2$, $\alpha_2=5$, $\gamma=0$ and $S=1$. }
\label{amor2}
\end{figure}

Figure \ref{amor2} illustrates the effect of the radiant cutoff for $I^*=0.003$ and the case described
above.  The radiant cutoff, as defined in Section 2.6, moves ${\rm d}N(v)$ to slightly larger 
values (Fig. \ref{amor2}a), and leads to narrower distributions of $a$, $e$ and $i$. More 
aggressive radiant selection criteria, such as the ones used in Campbell-Brown (2008) to define 
the helion/antihelion sources, would produce a slightly larger effect. On the other hand, Galligan 
\& Baggaley (2005) adopted a very broad radiant cutoff ($-120^\circ<l<-30^\circ$ and $20^\circ<l<120^\circ$, 
respectively, for our definition of $l$, and no condition on $b$). According to our tests, 
these broad selection criteria give results that are similar to those with no radiant cutoff.
 
\begin{figure}
\epsscale{0.9}
\plotone{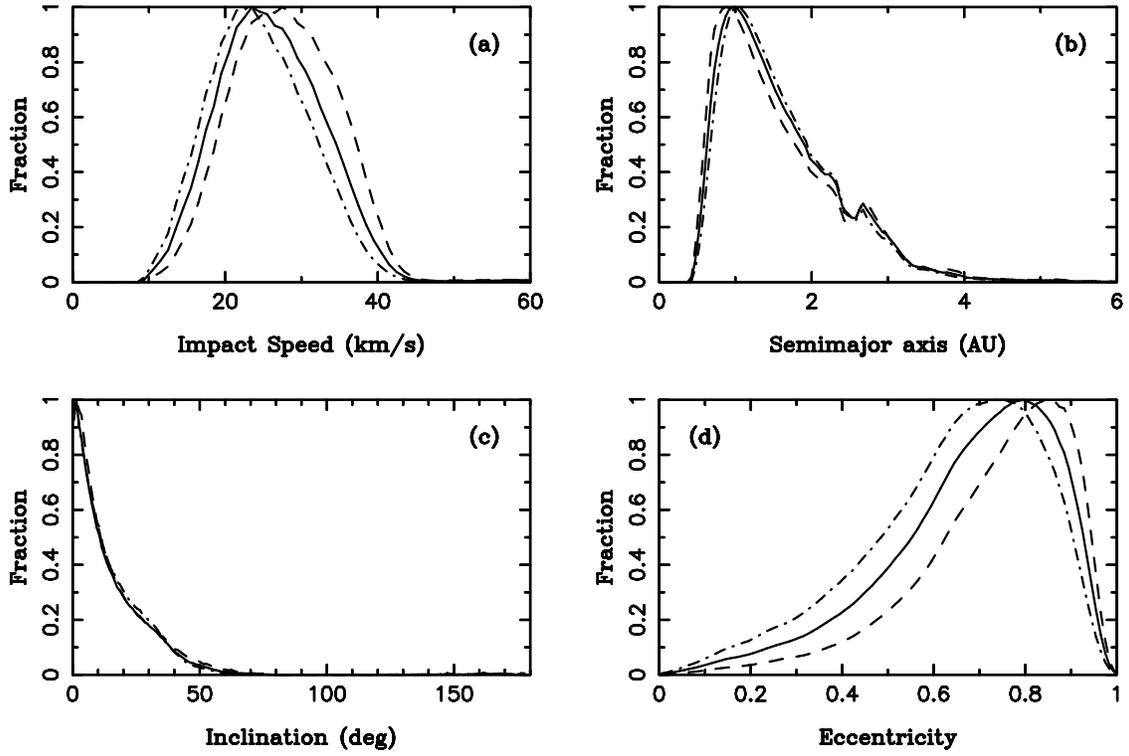}
\caption{Effect of $\gamma$. Different lines show the results for $\gamma=0$ (solid), 
$\gamma=-1$ (dashed), and $\gamma=1$ (dot-dashed). As in Fig. \ref{amor2}, we used 
$I^*=0.003$, $D^*=100$ $\mu$m, $\alpha_1=2$, $\alpha_2=5$ and $S=1$. No radiant cutoff 
was applied here.}
\label{amor3}
\end{figure}

The effect of the radial distribution of initial orbits is illustrated in Fig. \ref{amor3}. 
As expected, $\gamma<0$ produces ${\rm d}N(v)$ that peaks at larger values, and ${\rm d}N(e)$ 
that is more skewed toward $e=1$. This is because more particles are produced with small $q$ 
values in this case, and these particles tend to have larger $v$ and $e$ values when they impact. 
The effects of $\gamma>0$ are opposite to those of $\gamma<0$. Interestingly, ${\rm d}N(a)$ and 
${\rm d}N(i)$ obtained with $I^*=0.003$ are not very sensitive to changes of $\gamma$. 

Together with Fig. \ref{amor1}, these results show that it can be difficult to constrain 
the value of $\gamma$ from the fits to the AMOR measurements alone, because the effects of $\gamma$ are 
similar to those produced by slight changes in the detection efficiency, and can be confused with 
them. A detailed knowledge of the instrument sensitivity, that goes beyond the simple concept of the 
ionization cutoff described in Section 2.6, will be required for a more constrained 
modeling (see, e.g., Fentzke \& Janches 2008, Fentzke et al. 2009).  

\begin{figure}
\epsscale{0.9}
\plotone{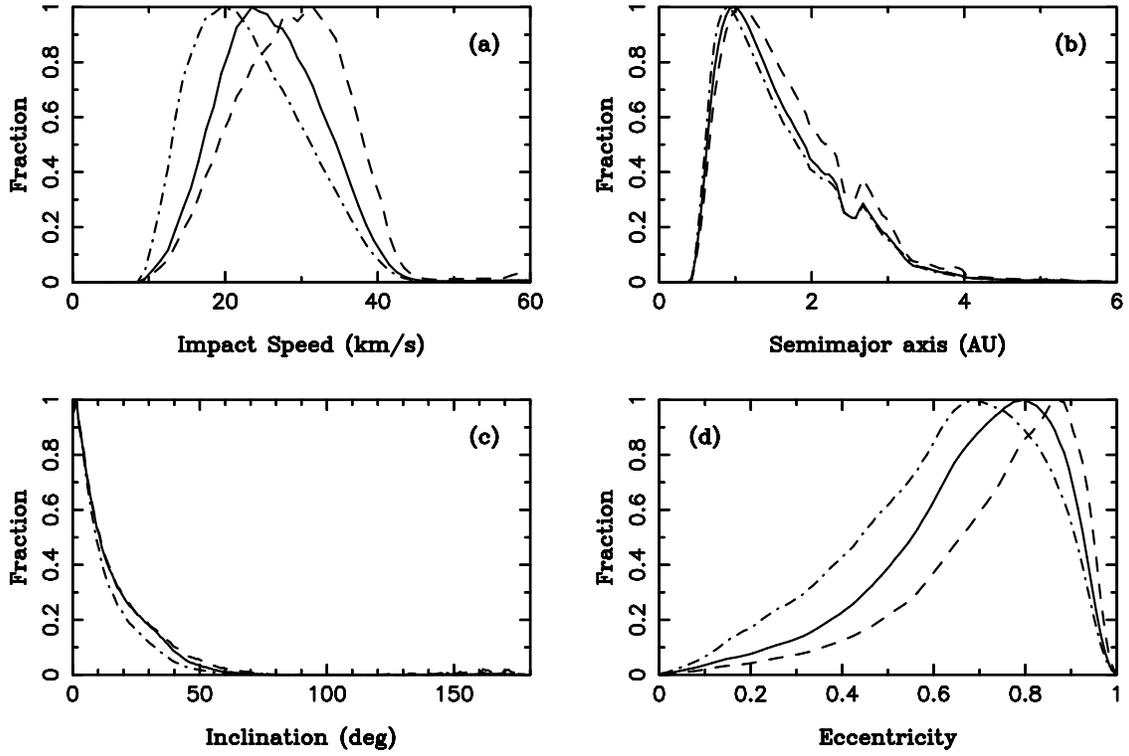}
\caption{Effect of $D^*$. Different lines show the results for $D^*=100$ $\mu$m (solid), 
$D^*=30$~$\mu$m (dashed), and $D^*=300$ $\mu$m (dot-dashed). As in Fig. \ref{amor2}, we used 
$I^*=0.003$, $\alpha_1=2$, $\alpha_2=5$ and $S=1$. No radiant cutoff was applied here.}
\label{amor4}
\end{figure}

The effects of initial ${\rm d}N(D)$ of particles, as discussed below, are in many ways 
similar to those produced by changes of $\gamma$ and $I^*$. Figure \ref{amor4} illustrates
the effect of $D^*$. Again, ${\rm d}N(v)$ and ${\rm d}N(e)$ show more variation than 
${\rm d}N(a)$ and ${\rm d}N(i)$. While for $D^*=30$~$\mu$m, the velocity peak shifts to 
$v\simeq30$ km s$^{-1}$, it moves to $v\simeq20$ km s$^{-1}$ for $D^*=300$ $\mu$m. This variation
can be linked to the ionization cutoff. For example, with $D^*=30$ $\mu$m, particles
tend to be smaller, and will be detected with $I^*=0.003$ only if their speeds are larger.

\begin{figure}
\epsscale{0.9}
\plotone{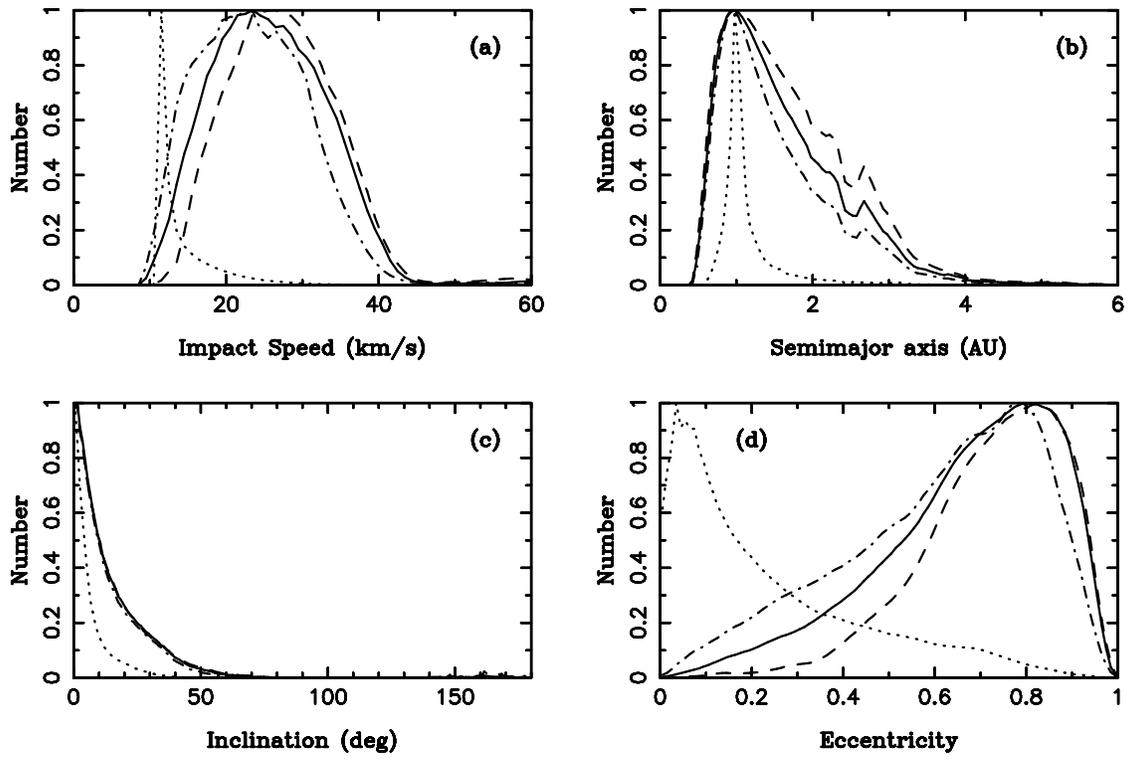}
\caption{The same as Fig. \ref{amor1} but for $\alpha_1=\alpha_2=3.5$.}
\label{amor5}
\end{figure}

The distributions ${\rm d}N(v)$ shown in Figs. \ref{amor1}a-\ref{amor4}a are all slightly
narrower than the one indicated by observations (cf. Fig. \ref{amor0}a). This difference 
cannot be resolved by varying $\gamma$, $S$ or $I^*$. Instead, to resolve this problem, we needed to 
assume that the power index of ${\rm d}N(D)$ is $3\lesssim\alpha\lesssim4$, at least in the 
size range relevant to AMOR observations.
To illustrate this case, Fig. \ref{amor5} shows the distributions for $\alpha=\alpha_1=\alpha_2=3.5$. 
While ${\rm d}N(a)$, ${\rm d}N(e)$ and ${\rm d}N(i)$ have not changed much relative to Fig. \ref{amor1}, 
the new distribution ${\rm d}N(v)$ with $I^*=0.01$-0.001 becomes broader, thus better mimicking the AMOR 
measurements. 

\begin{figure}
\epsscale{0.4}
\plotone{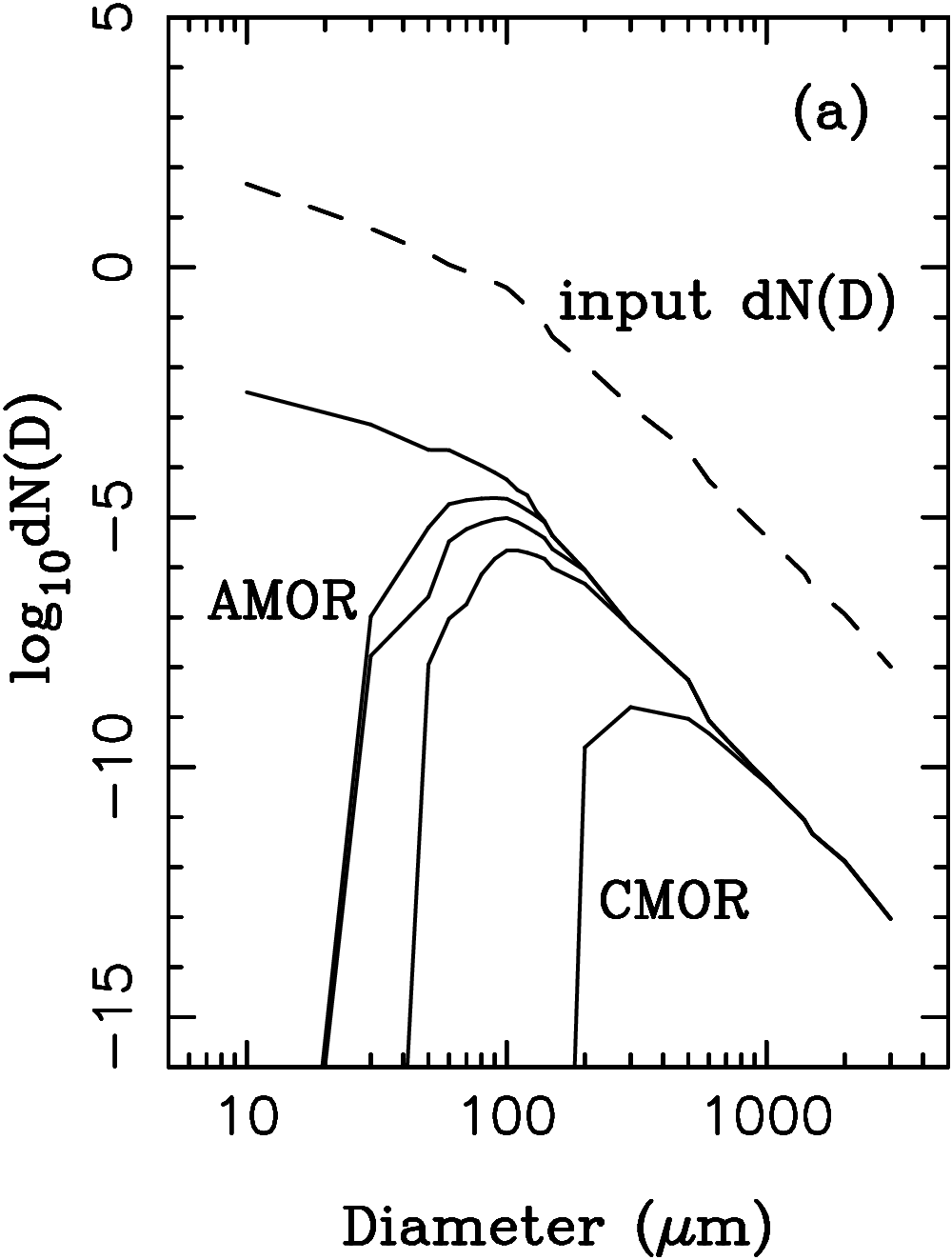}\hspace*{5.mm}
\plotone{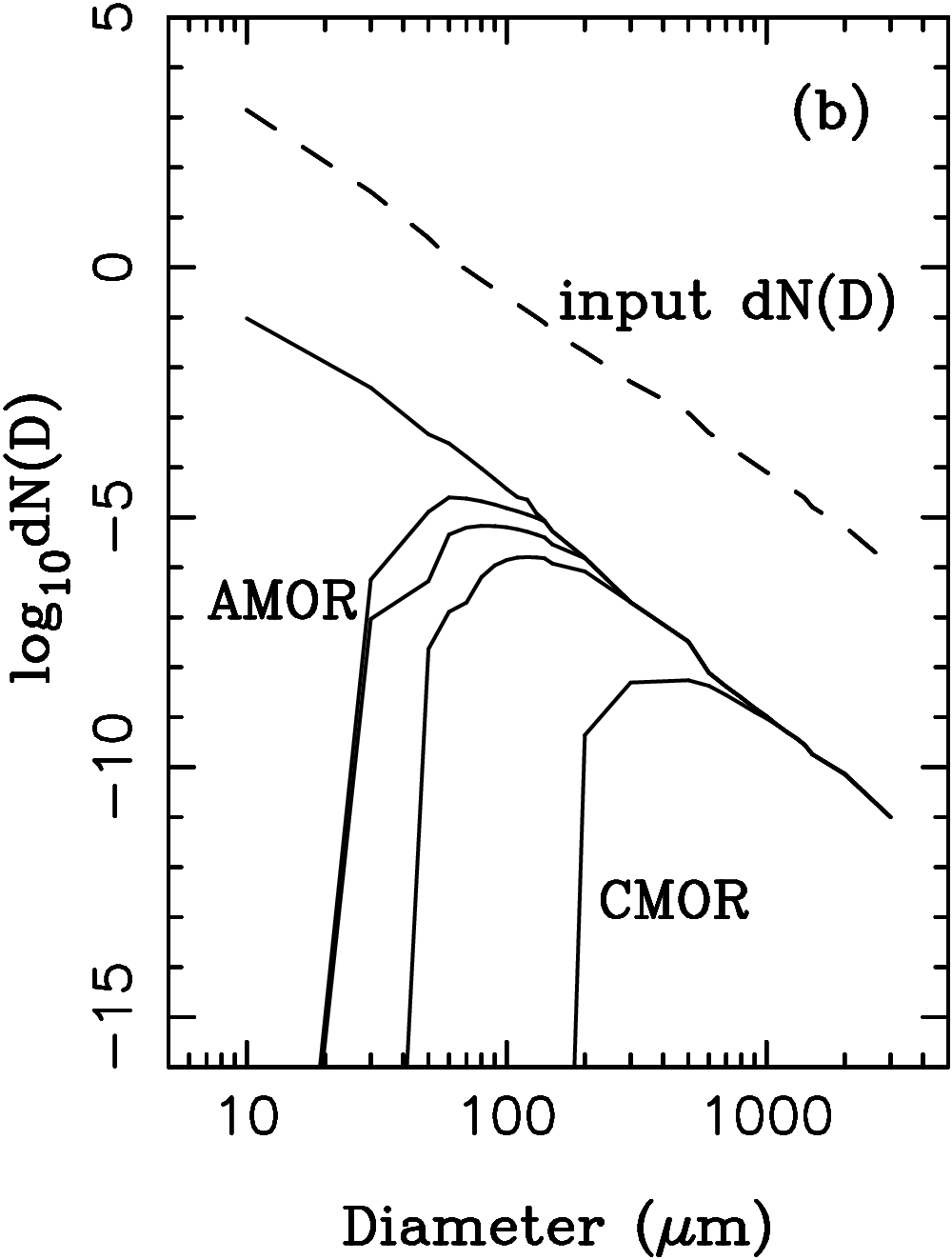}
\caption{The SFDs of different populations for: (a) $D^*=100$ $\mu$m, $\alpha_1=2$, $\alpha_2=5$, 
and (b) $\alpha_1=\alpha_2=3.5$. The input distributions, shown by dashed lines, correspond 
to those used in Figs. \ref{amor1} and \ref{amor5}. They were normalized to 1 particle with 
$D>100$ $\mu$m. The upper solid line in each panel shows the SFD of particles accreted by the
Earth ($I^*=0$). Accreted particles show a slightly steeper slope than the input SFD for $D\gtrsim100$ 
$\mu$m, because of the effects of disruptive collisions, which eliminate large particles in the 
G85 model, and a shallower slope for $D\lesssim100$ $\mu$m, because small JFC particles have 
smaller Earth-accretion probabilities than the larger ones due to their shorter P-R drag 
lifetimes. The other solid lines show the expected meteor SFD for AMOR (three lines corresponding,
from left to right, to $I^*=0.001$, 0.003 and 0.01) and CMOR ($I^*=1$).}
\label{sfd1}
\end{figure}

This trend can be easily understood. With a sharp SFD break at $D^*\sim100$ $\mu$m,
the particles that produce most meteors with $I>0.01$-0.001 are those with $D\sim100$ $\mu$m.
These particles have similar orbital histories and produce a relatively narrow ${\rm d}N(v)$.
With $\alpha_1=\alpha_2 \sim 3.5$, on the other hand, the size range of particles significantly 
contributing to AMOR meteors increases, relative to the previous case, producing a larger
variability in orbital histories, and thus also a larger spread in $v$. Figure \ref{sfd1} 
illustrates these trends. 

\begin{figure}
\epsscale{0.9}
\plotone{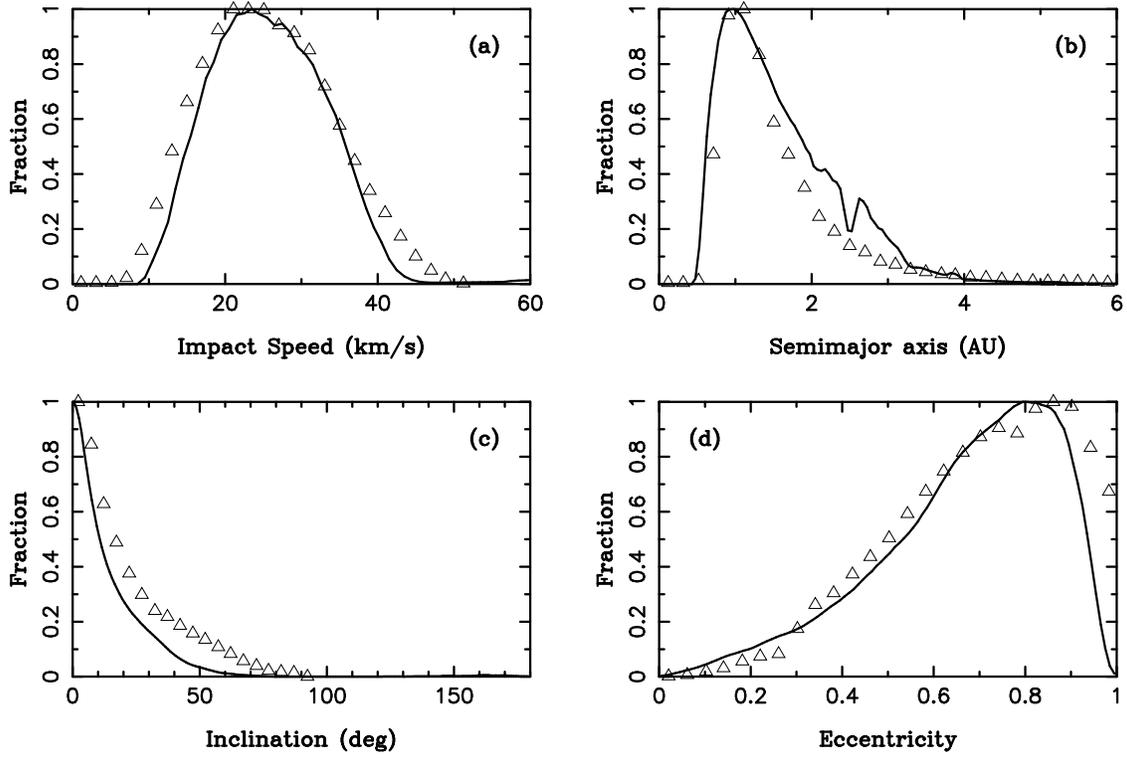}
\caption{Our preffered model for the AMOR meteors. The triangles label the corrected AMOR 
data from Galligan \& Baggaley (2005). Solid lines show our results obtained with $I^*=0.003$, 
$D^*=200$ $\mu$m, $\alpha_1=3.5$, $\alpha_2=5.0$, $\gamma=0$ and $S=1$.}
\label{amor7}
\end{figure}

By experimenting with different SFDs, we found that the best matches to AMOR observations 
can be obtained with $D^*\lesssim50$ $\mu$m and $\alpha_2=3.5$, or with $D^*\gtrsim200$ 
$\mu$m and $\alpha_1=3.5$, while the values of $\gamma$ and $S$ are essentially unconstrained 
(but see Section 3.2 for a discussion of the collisional model). We opt for $D^*=200$ $\mu$m
in Fig. \ref{amor7}, which illustrates one of our preferred models, because the original 
interpretation of spacecraft impact experiments indicates a change of slope at $D\simeq 
200$ $\mu$m (e.g., G85, Love \& Brownlee 1993). The solutions with $D^*\lesssim50$ $\mu$m and 
$\alpha_2=3.5$, which would better correspond to the reinterpretation of the impact experiments 
by Mathews et al. (2001), are also plausible. We will discuss this issue in Section 4.

While our model's ${\rm d}N(v)$, ${\rm d}N(a)$ and ${\rm d}N(e)$ in Fig. \ref{amor7} match observations
reasonably well, the model ${\rm d}N(i)$ is slightly narrower than the one measured by AMOR.
This indicates that we may be missing sources with larger inclinations. For example, as discussed 
in Section 2.1, our model for the initial inclinations of JFC particles can be inappropriate if JFCs 
lose mass gradually by recurrent splitting events (e.g., LD97, Di Sisto et al. 2009, N10). It is 
also possible, however, that the radiant cutoff of Galligan \& Baggaley (2004, 2005) is not 
sufficiently restrictive to pick up JFC meteoroids only, as hinted on by Fig. \ref{amor0}c, where 
${\rm d}N(i)$ seems to follow different trends for $i<30^\circ$ and $i>30^\circ$. Note that Campbell-Brown 
(2008), using a more restrictive radiant cutoff, obtained a relatively narrow ${\rm d}N(i)$ of 
helion/antihelion meteors.    

\begin{figure}
\epsscale{0.6}
\plotone{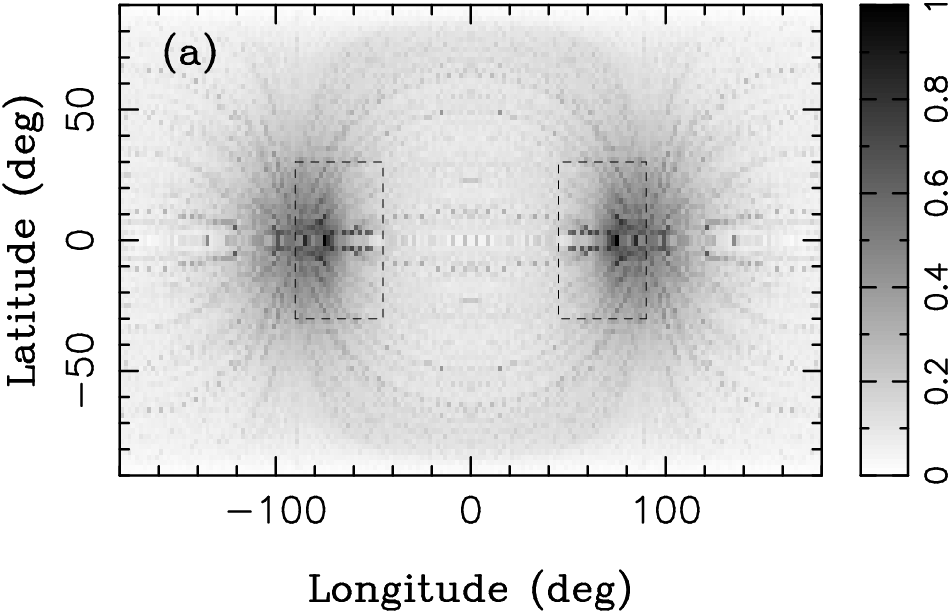}\\
\vspace*{0.5cm}
\plotone{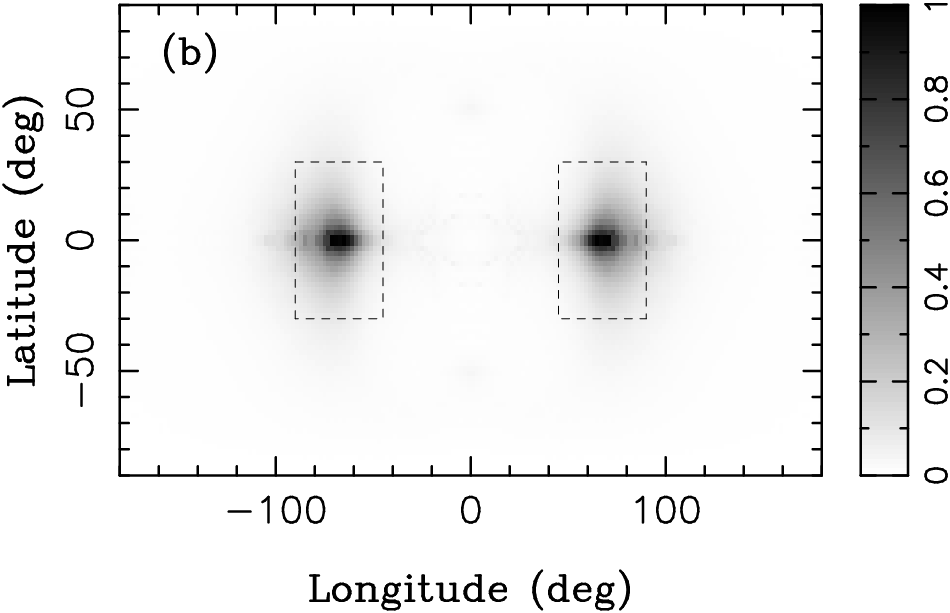}
\caption{Radiants for $I^*=0$ (a) and $I^*=0.003$ (b). The model parameters used here are the same 
as in Fig. \ref{amor7}. The gray scale shows the radiant density that was normalized to 1 at its 
maximum. The dashed rectangles in both panels show our radiant selection criteria
defined in Section 2.6.}
\label{amor_rad}
\end{figure}

Figure \ref{amor_rad} shows the radiant distributions for our preferred model shown in Fig.
\ref{amor7}. With $I^*=0$, the radiants fill the whole sky and show broad concentrations
around $l=-90^\circ$, $l=90^\circ$ and $b=0^\circ$. With $I^*=0.003$, however, the radiants become
tightly clustered about $l=-70^\circ$, $l=70^\circ$ and $b=0^\circ$. 
This highlights the importance of the ionization cutoff. For a comparison, Galligan \& Baggaley 
(2005) found that the helion and antihelion sources are centered at $l=\pm70^\circ$ and their full
widths are $\simeq20^\circ$ in both $l$ and $b$. The location and spread of our model radiants very 
closely match these measurements.

\subsection{CMOR}

\begin{figure}
\epsscale{0.9}
\plotone{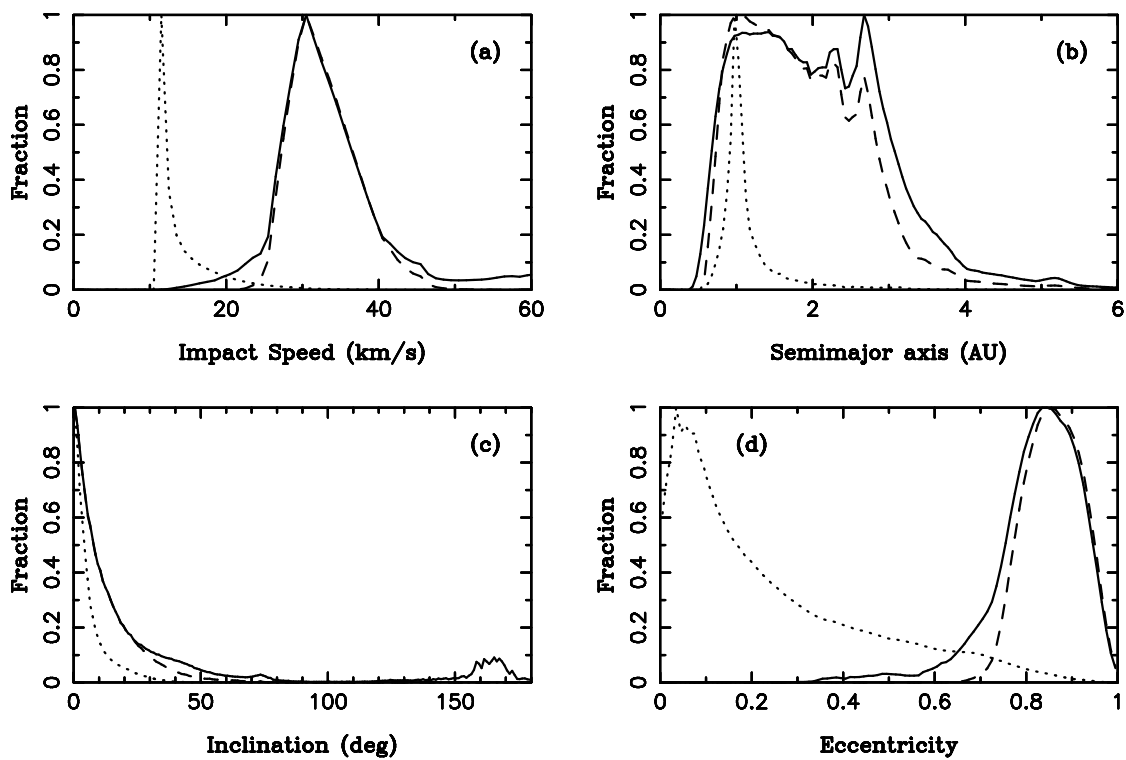} 
\caption{Different lines show the results for $I^*=0$ (dotted), $I^*=1$ (solid), and $I^*=1$ with radiant 
cutoff (dashed). As in Fig. \ref{amor7}, illustrating our preferred fit to the AMOR data, we used 
$D^*=200$ $\mu$m, $\alpha_1=3.5$, $\alpha_2=5$, $\gamma=0$ and $S=1$. The distributions shown here do 
not match the CMOR observations of the helion/antihelion meteors (cf. Figs. 10 and 12 in Campbell-Brown 2008).
They may be more similar to the distributions inferred from the visual meteor surveys that are sensitive to 
larger, $\sim$1 cm meteoroids (Jenniskens et al., in preparation).}
\label{cmor1}
\end{figure}

The results discussed in Section 3.1 were obtained with the standard G85 model for the collisional disruption 
of particles. In G85, the large, mm-sized particles have very short physical lifetimes ($\sim10^4$ yr) 
and disrupt before they can significantly evolve by P-R drag. Small particles, on the other hand, have 
long collisional lifetimes and evolve faster by P-R drag. The G85 model therefore implies 
different orbital histories of small and large particles and, as we found here, produces significantly 
different distributions of impact speeds and heliocentric orbits for $I^*=0.003$ and $I^*=1$. Figure 
\ref{cmor1} illustrates the case of $I^*=1$. These results are at odds with observations, because the 
distributions measured by AMOR and CMOR are not that different.

\begin{figure}
\epsscale{0.9}
\plotone{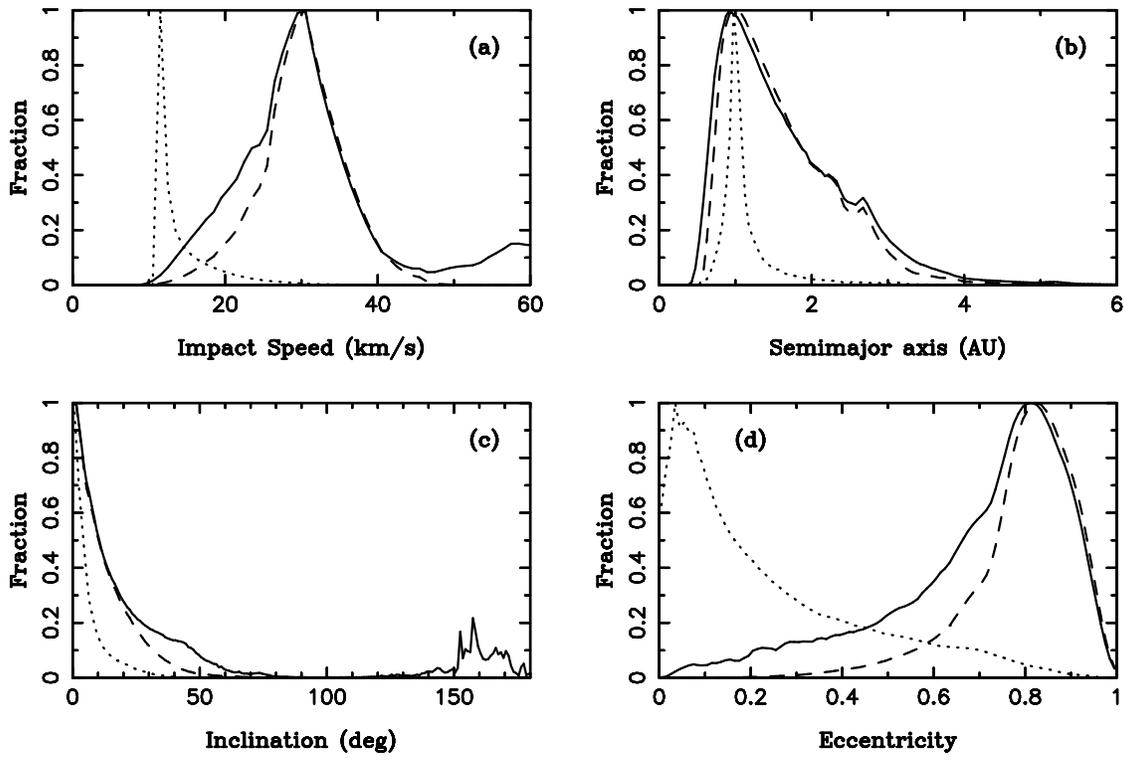} 
\caption{The same as Fig. \ref{cmor1} but for $S=100$.}
\label{cmor2}
\end{figure}

To resolve this problem, we needed to assume that $\tau_{\rm coll}$ for $D\sim1$ mm is significantly
longer than in G85 ($S \gtrsim 30$). Figure \ref{cmor2} shows the results for $S=100$. The main 
improvement with respect to Fig. \ref{cmor1} is that ${\rm d}N(a)$ now peaks at $a\sim1$ AU. This 
is a direct consequence of longer $\tau_{\rm coll}$, which allows the large particles to accumulate
larger P-R drifts, and reach $a\sim1$ AU. On the other hand, models with $S \gtrsim 30$ and 
$I^*=0.01$-0.001 do not match the AMOR measurements. This shows that the size dependence of the 
G85 collisional model may be incorrect. Taken together, if $S \gtrsim 30$ is needed to match CMOR, 
while $S \sim 1$ is needed to match AMOR, $\tau_{\rm coll}(D)$ should be more constant over the 
relevant size range, $D\sim30$-1000 $\mu$m according to Fig. \ref{sfd1}, than suggested by G85.

\begin{figure}
\epsscale{0.9}
\plotone{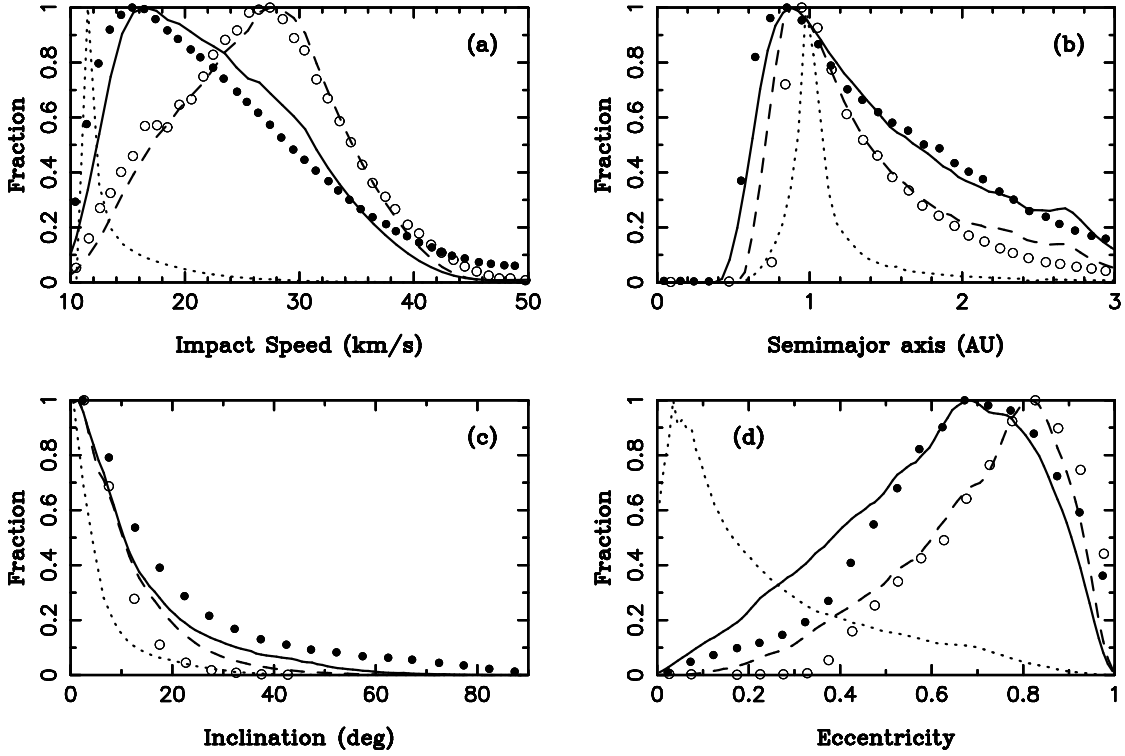} 
\caption{Our preferred model for the CMOR meteors. The filled and unfilled circles label the CMOR 
data from Figs. 10 and 12 in Campbell-Brown (2008). The filled circles are raw CMOR data where
no radiant cutoff was applied to separate different sources. The unfilled circles show 
the distributions, where the antihelion source was isolated by the radiant cutoff defined in Table 1
in Campbell-Brown (2008). These distributions were corrected for the observing biases and mass weighted.
The helion meteors, not shown here, have the corrected CMOR distributions very similar to those plotted 
here. Different lines show the model results for $I^*=0$ (dotted), $I^*=1$ (solid), and $I^*=1$ with 
radiant cutoff (dashed). We used the same parameters as in Fig. \ref{cmor1}, except for $S=100$ and 
$\alpha=2$. These assumed parameters are not inconsistent with those used is Fig. 10, because they apply 
to larger particles. Figures 10 and 14 can therefore be thought as a simultaneous fit to the radar data. 
Note that the X-axis ranges were changed here relative to the previous figures to show things more clearly.}
\label{cmor3}
\end{figure}

We performed a search in parameter space to see whether we can obtain the impact speed and orbit
distributions with $S=100$ that would closely resemble those measured by CMOR (Fig. 10-12 in 
Campbell-Brown 2008). We found that the model results with $\alpha \sim 2$ work best. 
With $\alpha \sim 2$, at least locally near $D\sim500$ $\mu$m, which are the most important sizes
for CMOR, the model distributions have the characteristic shapes measured by CMOR (Fig. \ref{cmor3}). 
When no radiant cutoff is used, ${\rm d}N(v)$ has the maximum just below 
$v=20$ km s$^{-1}$, and ${\rm d}N(e)$ peaks at $e\sim0.7$. When the radiant cutoff is applied, 
${\rm d}N(v)$ has the maximum just below $v=30$ km s$^{-1}$, and ${\rm d}N(e)$ peaks at $e\sim0.8$.
These trends correspond very well to those in Figs. 10 and 12 in Campbell-Brown (2008).

The model ${\rm d}N(a)$ with radiant cutoff becomes more tightly clustered at $a\sim1$ AU than in the case
without cutoff, in a nice correspondence to the CMOR measurements (Fig. \ref{cmor3}b). Our ${\rm d}N(i)$ 
with radiant cutoff is slightly broader than the CMOR distribution (Fig. \ref{cmor3}c), which is logical 
because the radiant cutoff used by Campbell-Brown (2008) is more restrictive than the one used here.
Overall, the agreement between the model and observations is very good.

While the meteoroid SFD can be wavy (e.g., Ceplecha et al. 1998), with $\alpha \sim 3.5$ at 
$D\sim100$ $\mu$m (see Section 3.1) and $\alpha \sim 2$ at $D\sim500$ $\mu$m, these slope determinations can also 
be artificially imposed on the results by our simple treatment of the radar's detection 
efficiency.\footnote{The SFD constraints established here were obtained with the simple parametrization 
of the radar's detection efficiency described in Section 2.6. In reality, the detection efficiency should be a more complex
function of the meteoroid mass and speed, other impact parameters, and observing conditions. It is 
plausible, for example, that the detection probability of a meteor, $D(I)$, goes smoothly from $\sim$0 
for $I\leq I^*_0<I^*$ to $\sim$1 for $I\geq I^*_1>I^*$, and attains some intermediate values from $I^*_0$ 
to $I^*_1$. If so, this could broaden the size range of particles that contribute to detections and 
potentially resolve the problem with the width of ${\rm d}N(v)$, without the need to resorting to 
a relatively shallow SFD slope.} In addition, it is not clear to us whether it is appropriate to compare our results
for the helion/antihelion meteors to Figs. 10 and 12 in Campbell-Brown (2008), because their Fig. 10 
shows the CMOR distributions for {\it all} sporadic sources, and the distributions in Fig. 12 were 
weighted to a constant limiting mass, a correction that is not needed for a comparison with our model. 

\begin{figure}
\epsscale{0.6}
\plotone{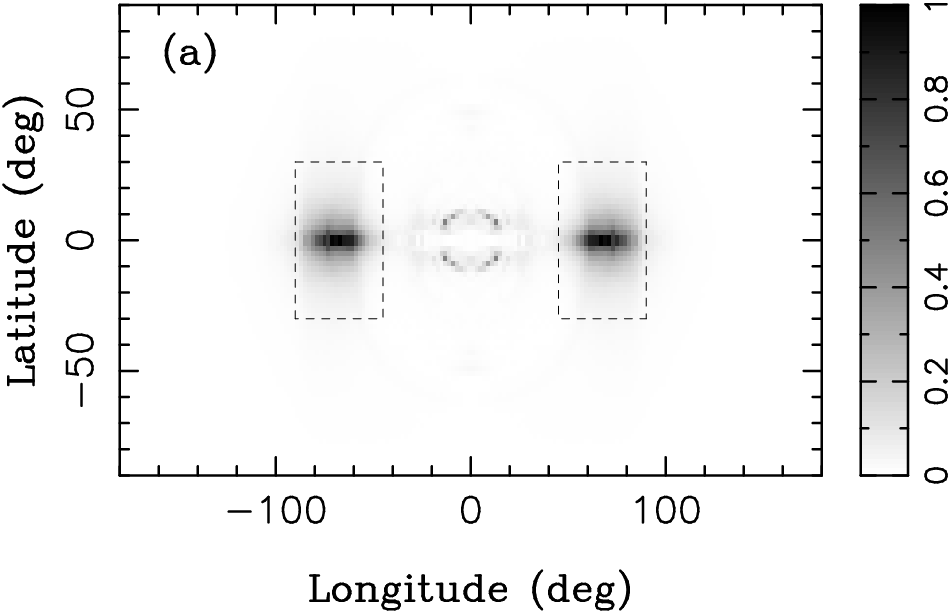}\\
\vspace*{0.5cm}
\plotone{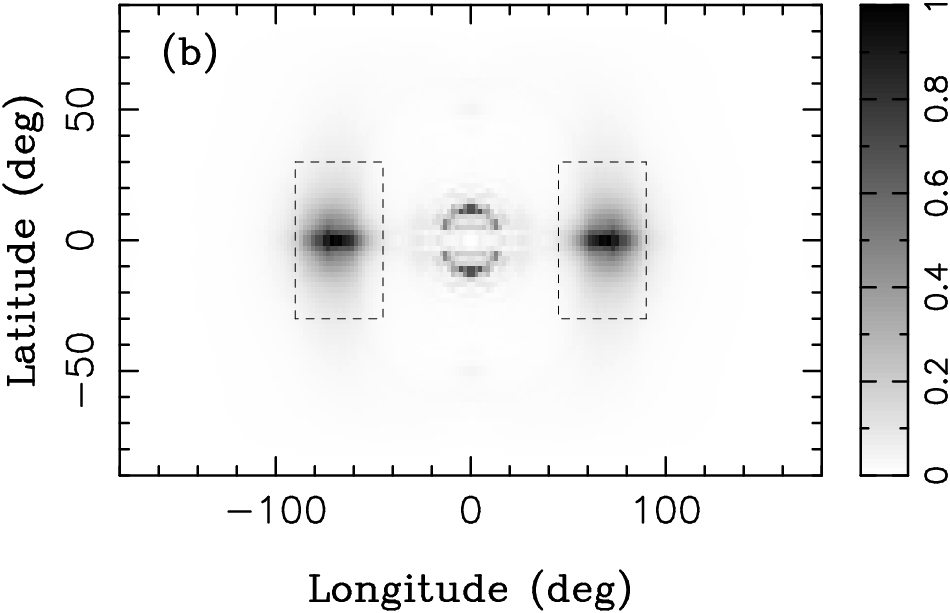}
\caption{Radiants for $I^*=1$ and: (1) $S=1$, and (b) $S=100$. The model parameters used here
are the same as in Figs. \ref{cmor1} and \ref{cmor2}. The gray scale shows the radiant 
density that was normalized to 1 at its maximum. The dashed rectangles in both panels 
denote our radiant selection criteria defined in Section 2.6.}
\label{cmor_rad}
\end{figure}

Finally, Fig. \ref{cmor_rad} shows the CMOR radiant distributions for models illustrated in Figs. 
\ref{cmor1} and \ref{cmor2}. With longer $\tau_{\rm coll}$, the radiants are slightly more spread   
around the helion and antihelion directions. Interestingly,  Fig. \ref{cmor_rad} indicates that the 
JFC meteoroids are capable of producing apex meteors. These apex particles have prograde orbits,
low impact speeds, and very low semimajor axes. They impact from the apex direction because their 
orbital speed near the aphelion at 1 AU is smaller that the Earth's orbital speed. The contribution 
of JFC meteoroids to apex meteors should be small, however, relative to those produced by the retrograde 
HTC and/or OCC meteoroids. We verified that only a small fraction ($<$1\%) of the JFC meteoroids can 
reach retrograde orbits.  
\subsection{IRAS}
Using the methods described in Sections 2.4 and 2.5, we computed the MIR fluxes for all
models considered in the previous section. Here we illustrate these results and compare them with
with those obtained by IRAS. Before we do so, however, we want to emphasize that the ZC is in all 
likelihood a mixture of several particle populations, including contribution from asteroids and 
long-period comets (see, e.g., N10), while here we only model the JFC component. The best fits obtained 
to IRAS observations in this work are therefore only approximate, and could be modified if other 
components of the ZC were considered. 

Figure \ref{sirt1} shows our results for $D^*=100$ $\mu$m, $\alpha_1=2$, $\alpha_2=5$, $\gamma=0$ and 
$S=1$, corresponding to Figs. \ref{amor1} and \ref{amor2}. The model MIR profiles are slightly narrower 
than the observed ones, but otherwise correspond to IRAS measurements reasonably well. For example, 
a small, $\lesssim10$\% contribution from a source with a more isotropic distribution of inclinations,
such as HTCs and/or OCCs, would easily compensate for the small difference. See N10 for a discussion
of additional sources that were not modeled here.   
  
A different way to bring the model and observations into a closer agreement is to assume that $\gamma<0$. 
With $\gamma<0$, the distribution of JFC particles is weighted toward low $R$, and is projected to a wider 
range of $b$ when observed from $R=1$ AU. If $\gamma>0$, on the other hand, the distribution is weighted
toward large $R$, and is seen closer to the ecliptic. We tested a continuous range of $\gamma$ values 
and found that $\gamma\simeq-1.3$ provided the best match to the IRAS observations (Fig. \ref{sirt2}). 

The effects of additional sources and $\gamma\neq0$ on the MIR profiles are to some degree degenerate 
in the IRAS model. They would be difficult to separate, based solely on modeling of the IRAS observations, 
if we included additional sources in the present work. For example, as discussed above, the MIR profiles 
become broader, and more similar to the IRAS measurements, if $\gamma<0$ and/or if sources with a more 
isotropic inclination distribution are included (N10). On the other hand, a small asteroid contribution 
at the $\sim$5-10\% level (Nesvorn\'y et al. 2006) would produce slightly narrower profiles than those 
obtained with the JFCs alone (N10). A two-source model with JFCs and asteroids would thus require $\gamma<0$.

The MIR profiles obtained in our model are not overly sensitive to the assumptions on the collisional 
lifetimes of particles. For example, increasing the collisional lifetime of mm-sized particles relative to 
the G85 model, which may be needed to match CMOR observations (Section 3.2), does not affect the results 
obtained here, because the ZC's cross section is mainly in $D\lesssim200$ $\mu$m particles (N10). 

In addition, the model profiles are also insensitive to the input SFD of the JFC particles. This is because 
the JFC particles with $D \lesssim 200$ $\mu$m have $\tau_{\rm coll}$ that exceeds their P-R drag 
lifetimes. All these particles therefore have roughly similar orbital histories and produce similar
MIR profiles. This explains why we obtain nearly identical results for all $D^*<200$~$\mu$m.\footnote{We 
note that the model profiles obtained for $\gamma\simeq-1.3$ and $D^*=10$ $\mu$m are slightly narrower 
than those shown in Fig. \ref{sirt2} for $D^*=100$ $\mu$m, because small particles drift faster and their 
inclinations are disturbed to a lesser degree by planets and planetary resonances.} The effects of 
$\alpha_1$ and $\alpha_2$ are also minor. 

\subsection{ZC Mass and Cross Section, and Mass Influx on Earth}

The comparison of our model with the IRAS data is important because it allows us to obtain the 
absolute calibration of the number of particles in the ZC (or, more precisely, their 
total cross-section). This calibration can then be used to estimate
the rate of the terrestrial accretion of interplanetary material, both with and without the 
ionization cutoff, with the former estimate being relevant to radar observations.
Note that it is more difficult to derive the overall terrestrial accretion rate from
the meteor radar measurements alone, because different radar instruments have different 
detection sensitivities, and some, such as the less sensitive CMOR, do not detect the very 
small and/or slow meteoroids (see discussion in Section 4). 

Unless we specify otherwise, all estimates quoted below were obtained for the full size range 
of particles between $D=5$ $\mu$m and $D=1$ cm ($10^{-10}$ g to 1 g for $\rho=2$ g cm$^{-3}$). 
These estimates need to be considered with caution because they were obtained with approximate 
initial SFDs. In reality, the number of particles released by JFCs can be a complicated 
function of $D$, and should also depend on the circumstances of the splitting/disruption events.

We start by discussing the total cross-section area ($\sigma_{\rm ZC}$) and mass ($m_{\rm ZC}$) of 
particles in the ZC. For the sake of simplicity, we will assume that $30\lesssim D^*\lesssim300$ $\mu$m,
$\alpha_1<3$ and $\alpha_2>4$, so that particles with sizes below 30 $\mu$m and above
300 $\mu$m do not strongly contribute to the cross section or mass, as suggested by 
the spectral observations of the ZC (e.g., Reach et al. 2003), and various other measurements (see, e.g., 
Ceplecha et al. (1998), and the references therein). 

\begin{figure}
\epsscale{0.8}
\plotone{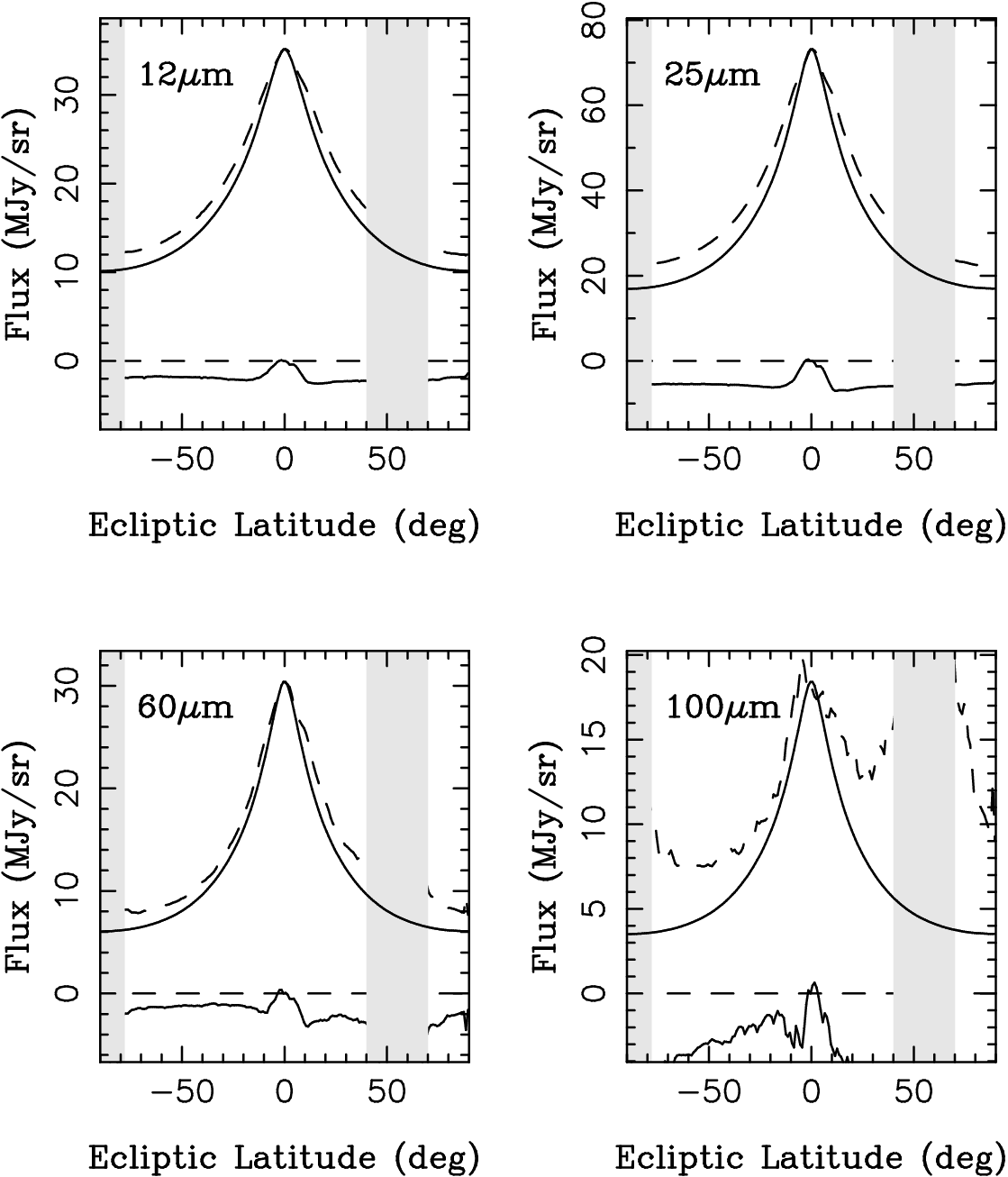}
\caption{The MIR profiles at: (a) 12 $\mu$m, (b) 25 $\mu$m, (c) 60 $\mu$m, and (d) 100 $\mu$m 
wavelengths. The dashed line shows the mean IRAS profiles for $l_\odot=90^\circ$. The upper solid curves 
show the model results for the same wavelength and elongation. The bottom lines show the residual 
flux obtained by subtracting the model flux from the mean IRAS profile. Here we used the same 
model parameters as in Fig. \ref{amor1}: $D^*=100$ $\mu$m, $\alpha_1=2$, $\alpha_2=5$, $\gamma=0$ and 
$S=1$.}
\label{sirt1}
\end{figure}

\begin{figure}
\epsscale{0.8}
\plotone{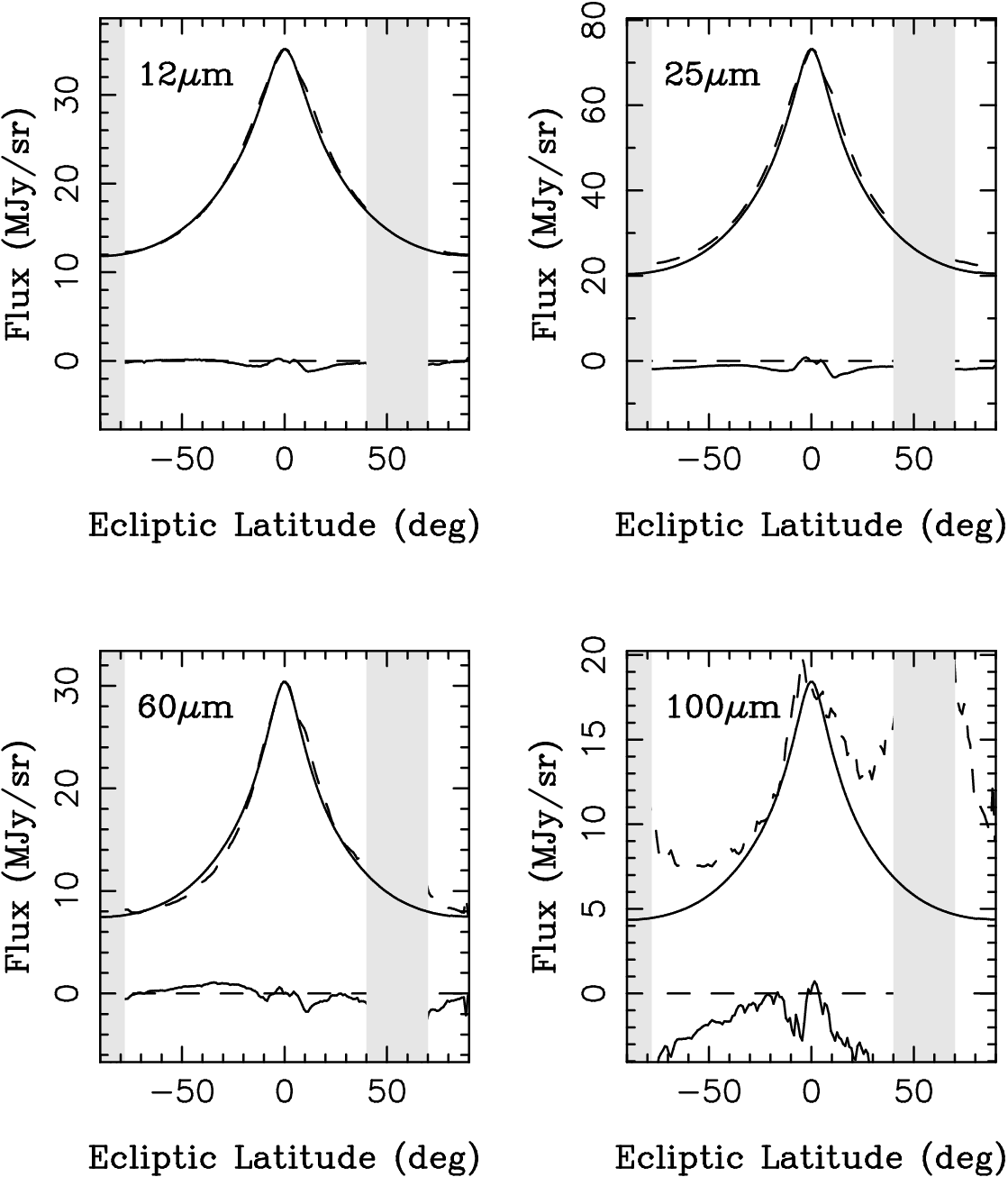}
\caption{The same as Fig. \ref{sirt1}, but with $\gamma=-1.3$. The radial distribution obtained with 
$\gamma=-1.3$ leads to the best fit to IRAS observations, at least for the input SFD assumed here. The 
MIR profiles, however, are not overly sensitive to the assumed SFD.}
\label{sirt2}
\end{figure}

With these assumptions we find that $1.7\times10^{11}<\sigma_{\rm ZC}
<3.5\times10^{11}$~km$^2$, where the larger values correspond to $D^*=300$ $\mu$m. This estimate
is in a good agreement with N10 who found that $\sigma_{\rm ZC}=(2.0\pm0.5) \times 10^{11}$ km$^2$. 
The uncertainty in $\sigma_{\rm ZC}$ mainly stems from the uncertainty in $D^*$, with $\gamma$ 
producing only a minor effect. For a reference, the models shown in Fig. \ref{sirt1} and \ref{sirt2} 
have $\sigma_{\rm ZC}=2.1\times 10^{11}$ and $2.0\times 10^{11}$ km$^2$, respectively (Table 1).

Mass $m_{\rm ZC}$ is more poorly constrained then $\sigma_{\rm ZC}$. For $30\lesssim D^*\lesssim300$ 
$\mu$m, $\alpha_1<3$, $\alpha_2>4$ and $\rho=2$ g cm$^{-3}$ we estimate that $10^{19}<m_{\rm ZC}
<1.5\times10^{20}$ g, with larger values corresponding to $D^*=300$ $\mu$m. For a more restrictive 
assumption with $D^*\simeq100$ $\mu$m, we find that $3\times10^{19}<m_{\rm ZC} <5\times10^{19}$ g,
where the exact value depends on $\alpha_1$, $\alpha_2$ and $\gamma$. For a reference, the 
models shown on Fig. \ref{sirt1} and \ref{sirt2} have $m_{\rm ZC}=3.8\times10^{19}$ and $3.9\times10^{19}$~g, 
which roughly corresponds to a 33-km-diameter sphere with $\rho=2$ g cm$^{-3}$.

These results compare well with those reported in N10, where it was found that $2.6\times10^{19}<m_{\rm ZC} 
<5.2\times10^{19}$ g, under the assumption that the continuous SFD can be approximated by a population 
of the same-size particles with $D=100$-200 $\mu$m.

N10 estimated that the input mass rate of $\dot{m}_{\rm ZC}=1,000$-1,500 kg s$^{-1}$ is needed to 
keep the ZC in a steady state. Here we obtain larger values, mainly because the population of particles 
released with low $q$ has shorter lifetimes and needs to be resupplied at a higher rate. If 
$D^*\lesssim100$ $\mu$m, $\dot{m}_{\rm ZC}$ ranges between 3,000 and 7,000 kg s$^{-1}$, with the largest
values corresponding to $\gamma=-1.3$ in the model illustrated in Fig. \ref{sirt2}. Input rate
$\dot{m}_{\rm ZC}$ is also sensitive to $D^*$, roughly in the same proportion as $m_{\rm ZC}$.
For example, $\dot{m}_{\rm ZC}\sim1,600$ and 19,000 kg s$^{-1}$ for $D^*=30$ $\mu$m and $D^*=300$ $\mu$m, 
respectively. These estimates are valid for $\alpha_1<3$ and $\alpha_2>4$. The required input rates 
can be somewhat smaller or larger if $\alpha_1 \sim 3.5$ and/or $\alpha_2 \sim 3.5$ (see Table 1).

Finally, we consider the terrestrial accretion rate, $\dot{m}_{I^*}$, where $\dot{m}_{I^*}$ denotes
the rate for $I>I^*$. We consider cases with $I^*=0$,  $I^*=0.003$ and $I^*=1$, with the latter two 
roughly corresponding to our expectations for AMOR and CMOR, respectively. Using the IRAS calibration, 
we find that our standard model with $D^*\simeq100$ $\mu$m, $\alpha_1<3$, $\alpha_2>4$ implies that 
$\dot{m}_0=(15,000\pm3,000)$ tons yr$^{-1}$, $\dot{m}_{0.003}=(5,000\pm2,000)$ tons yr$^{-1}$, and 
$\dot{m}_1=(500\pm400)$ tons yr$^{-1}$, where a large part of the quoted uncertainty comes from the poorly
constrained  $\gamma$. Note that the values for $I^*=0.003$ and $I^*=1$ do not include any radiant cutoff.

\begin{figure}
\epsscale{0.7}
\plotone{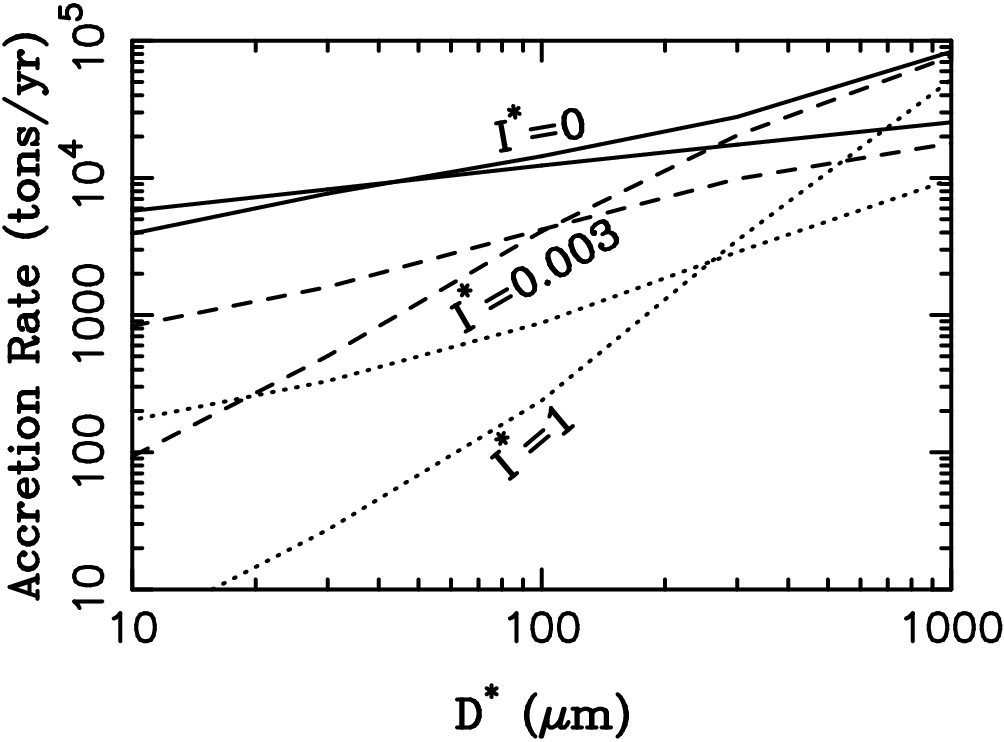}
\caption{The terrestrial accretion rate of JFC particles as a function of $D^*$ obtained in our model 
for $\gamma=0$ and $S=1$. Different lines denote the results for: $I^*=0$ (solid), $I^*=0.003$ (dashed), and $I^*=1$ 
(dotted). The two lines for each $I^*$ were computed for different values of $\alpha_1$ and $\alpha_2$. 
The more horizontal lines correspond to $\alpha_1=3$ and $\alpha_2=4$. The more inclined lines correspond
to $\alpha_1=2$ and $\alpha_2=5$. For $D^*\simeq100$ $\mu$m, the overall terrestrial accretion rate
for $I^*=0$ is $\sim1$-$2\times10^4$ tons yr$^{-1}$.}
\label{summary}
\end{figure}

The uncertainty becomes larger if $D^*$ is allowed to vary (Fig. \ref{summary}). For example, with 
$\alpha_1=2$, $\alpha_2=5$ and $\gamma=0$, we obtain $\dot{m}_0=26,000$ tons yr$^{-1}$, $\dot{m}_{0.003}=19,000$ 
tons yr$^{-1}$, and $\dot{m}_1=3,300$ tons yr$^{-1}$ for $D^*=300$ $\mu$m, and $\dot{m}_0=7,700$ 
tons yr$^{-1}$, $\dot{m}_{0.003}=500$ tons yr$^{-1}$, and $\dot{m}_1=27$ tons yr$^{-1}$ for $D^*=30$ 
$\mu$m. Also, our preferred model for the AMOR meteors illustrated in Fig. \ref{amor7} gives $\dot{m}_0=
12,000$ tons yr$^{-1}$, $\dot{m}_{0.003}=2,900$ tons yr$^{-1}$, and $\dot{m}_1=230$ tons yr$^{-1}$.
Similarly, the model with $D^*=50$ $\mu$m, $\alpha_1=2$ and $\alpha_2=3.5$ gives $\dot{m}_0=18,000$ 
tons yr$^{-1}$, $\dot{m}_{0.003}=9,300$ tons yr$^{-1}$, and $\dot{m}_1=4,100$, or
$\dot{m}_0=17,000$ tons yr$^{-1}$, $\dot{m}_{0.003}=8,500$ tons yr$^{-1}$, and $\dot{m}_1=3,300$,
tons yr$^{-1}$, if the size range of the contributing particles is restricted to $10<D<3000$ $\mu$m
(Table 1).

The above estimates with $I^*=0$ are a factor of several lower than those found by N10. This difference
probably stems from some of the crude approximations used by N10. For example, N10 did not use a continuous 
SFD of particles and approximated ${\rm d}N(D)$ by delta functions. Their initial particle orbits 
had (almost exclusively) $q>1.5$ AU, and did not take into account the fact that many JFCs can split 
and/or disrupt with $q<1.5$ AU. Moreover, N10 did not properly include the collisional lifetimes of 
JFC particles in their model. The results presented here, which include all these components, and which 
were validated on meteor observations, can be more trusted and should supersede those reported in N10. 
\section{Discussion}
The results reported in Section 3.4 show that the low-sensitive meteor radars such as CMOR can only detect  
a few percent of the overall mass flux. It may therefore be difficult to estimate the terrestrial 
accretion rate from these measurements alone. The more sensitive meteor radars such as AMOR, on the other 
hand, should detect 10-50\% of the flux, with the exact value mainly depending on the SFD assumptions (Table 
1). These more sensitive measurements, especially those obtained with the HPLA radars, are therefore better 
suited for estimating the overall accretion rate. 

The terrestrial accretion rate found here is comparable to that originally inferred by Love \& 
Brownlee (1993) from the LDEF experiment, and much larger than the one suggested by Mathews et al. (2001) 
from the Arecibo Observatory (AO) measurements of meteor fluxes. As pointed out by Mathews et al. 
(2001), the difference between the AO and LDEF measurements could be resolved if LDEF data were 
recalibrated to $v\sim50$ km s$^{-1}$, which is the prevailing meteor speed as seen at Arecibo (see 
Janches et al. 2003, 2006, Fentzke et al. 2009).  

We found that ${\rm d}N(v)$ for $I^*=0$ peaks at $v \sim v_{\rm esc}=11.2$ km s$^{-1}$ (see, e.g., Figs. 
\ref{amor1}, \ref{amor5} and \ref{amor7}). The only parameter choices that we were able to identify, 
where this was not the case, were those where it was assumed that essentially all dust was produced 
with $q\lesssim0.5$~AU, and that $D^*\gtrsim100$ $\mu$m for $S=1$. Such an extreme $q$ dependence seems 
unlikely, because not many solar system objects --potential parent bodies of meteoroids-- ever reach 
$q\lesssim0.5$~AU. [Comet 2P/Encke has $q=0.34$ AU, but as we discussed in the Section 1, the measured 
mass loss in comet 2P/Encke is far too low to be dominant.]
 
The case described above could potentially be interesting, because it could help to explain why 
the meteor observations at AO, albeit being much more sensitive than AMOR, do not detect a significant 
population of meteors with $v<15$ km s$^{-1}$ (e.g., Janches et al. 2008). To allow for $D^*<100$ $\mu$m in 
this case, and bring our results to a closer agreement with Mathews et al. (2001), 
$\tau_{\rm coll}$ of $D<100$-$\mu$m particles, mainly for orbits with $q\lesssim0.5$ AU, would need to be 
significantly shorter than suggested by G85. 

To match the CMOR measurements in our model, the collisional lifetime of meteoroids with $D\sim1$ mm needs to 
be significantly longer than suggested by G85. Such a long lifetime, of order of a few times $10^5$ yr for 
a circular orbit at 1 AU, can be difficult to reconcile with the inferred lifetimes of meteor streams 
that seem to disappear on a much shorter timescale ($<$few thousand years; e.g., Jenniskens 2008). Possibly,
the cm-sized particles released from JFCs, which appear to be dominant in the visual observations of the 
meteor streams, are physically weak and disrupt in a few thousand years. They could produce a population of 
mm-sized and smaller particles that, according to our work, could be more resistant to collisions. 

As explained in section 2.3, our model neglects small fragments produced by disruptions of larger 
particles, because it is difficult to account for numerous debris particles in the $N$-body code. Since 
the fragments are small, and should be released on the orbits already evolved by P-R drag, we may speculate 
that this could lead to a steeper SFD of particles with low perihelion distances. To compensate for that, 
our preferred model for the source population of particles would need to be adjusted. It is unclear, however, 
if the effect of the collisional cascade is important. Future work will need to address this problem.  
\section{Conclusions}
The radar observations of sporadic meteors reveal an important population of meteoroids that impact Earth 
from the helion and antihelion directions. Typically, these particles have heliocentric orbits with $a \sim 
1$ AU, $e>0.3$, $i<30^\circ$, and dive into the upper atmosphere at speeds $v \sim 20$-30 km s$^{-1}$. These 
results were seemingly inconsistent with the model of the circumsolar meteoroid complex developed in N10, which 
has been calibrated on the IRAS's MIR observations of the ZC's thermal emission. 

The N10 model implied that particles impacting Earth from the helion/antihelion directions should 
either have $a \sim 1$ AU and $e\lesssim0.3$, or $a \sim 2$-4 AU and $e \gtrsim 0.6$. The former case 
corresponds to $D\lesssim100$-$\mu$m meteoroids, whose orbits evolved by P-R drag. The latter case are 
the large, $D\gtrsim1$-mm particles that were assumed in N10 to be collisionally disrupted before 
their orbits could significantly evolve by P-R drag. The different orbital histories of small and large 
meteoroids in the N10 model would mean that the meteor radars with different detection thresholds
should measure very different distributions of the impact speeds and orbits. This is not the case.  
 
Here we showed that the above problem can be resolved if: (1) the N10 model is modified to account for the
detection efficiency of meteor radars; (2) meteoroids are released from JFCs over a range of perihelion 
distances with at least some fraction initially having $q \lesssim 1$ AU; and (3) $D\sim1$ mm particles 
have significantly longer ($\gtrsim$30 times) collisional lifetimes than those estimated in G85. 
With these modifications of the N10 model, the results match meteor constraints.

We also found, using the AMOR observations as a constraint, that $D\sim100$-$\mu$m particles cannot have 
much longer collisional lifetimes than proposed in G85. Together with (3), these results therefore suggest that 
$D\sim100$ $\mu$m and $D\sim1$ mm meteoroids may have more comparable collisional lifetimes (a few times $10^5$ yr for a 
circular orbit at 1 AU) than thought before. If so, the SFD shape inferred from the measurements of the 
spacecraft impact detectors (such as, e.g., Ulysses, Galileo, LDEF; G85, Love \& Brownlee 1993) 
may be more closely related to the initial SFD of particles released at sources then to the collisional 
destruction of particles in space.  

We showed that the modified N10 model can successfully match the telescopic observations of the zodiacal 
cloud. Using IRAS to calibrate the model, we estimated that the cross section and mass of the zodiacal cloud 
are $\sigma_{\rm ZC}= (1.7$-$3.5)\times10^{11}$ km$^2$ and $m_{\rm ZC}\sim 4\times10^{19}$~g. The terrestrial 
accretion rate of JFC particles was found to be $\sim15,000$ tons yr$^{-1}$, of which only a few percent 
should be detected by CMOR, and 10-50\% should be detected by the more sensitive AMOR. 

Moreover, some $10^3$-$10^4$ kg s$^{-1}$ of material must be provided by JFCs to keep the ZC in a 
steady state. This new input mass estimate is up to $\sim$10 times larger than the one suggested by 
N10 (see also Leinert et al. 1983), because particles starting with low $q$ have shorter lifetimes, and 
need to be resupplied at a faster rate. This new estimate resonates with the N10 model in which the ZC is 
dominated by the meteoroids released by disrupting/splitting JFCs, because the observed activity of JFCs 
cannot provide the needed input. 
\acknowledgements
This article is based on work supported by the NASA's Planetary Geology and Geophysics and
Planetary Astronomy programs. DJ's participation in this work
was supported through the NSF Award AST 0908118. The work of DV was partially
supported by the Czech Grant Agency (grant 205/08/0064) and the Research Program MSM0021620860
of the Czech Ministry of Education. We thank Margaret Campbell-Brown, Jack Baggaley and Hal Levison
for useful discussion. We also thank the anonymous reviewer for excellent suggestions to this 
work.

\clearpage

\begin{table}[t]
\begin{center}
\begin{tabular}{rrrrrrrrrr}
\hline 
$D^*$     & $\alpha_1$ & $\alpha_2$ & $\gamma$ & $\sigma_{\rm ZC}$ & $m_{\rm ZC}$ & $\dot{m}_{\rm ZC}$ &  $\dot{m}_0$ & $\dot{m}_{0.003}$ & $\dot{m}_1$ \\
$\mu$m    &            &            &          & $10^{11}$ km$^2$ & $10^{19}$ g & kg s$^{-1}$   & tons yr$^{-1}$   &  tons yr$^{-1}$   & tons yr$^{-1}$   \\
\hline 
100 & 2  & 5 & 0       & 2.1    & 3.8        & 4,200              & 15,000      & 4,200         &  240         \\
100 & 2.9 & 4.1 & 0    & 2.1    & 4.6        & 6,200              & 12,000      & 4,100         &  860         \\ 
100 & 2  & 5 & -1      & 2.0    & 3.8        & 5,200              & 15,000      & 6,100         &  480         \\                  
100 & 2  & 5 & -1.3    & 2.0    & 3.9        & 5,800              & 16,000      & 7,000         &  590         \\ 
100 & 2  & 5 & 1       & 2.3    & 4.0        & 4,000              & 14,000      & 3,100         &  130         \\ 
30  & 2  & 5 & 0       & 1.8    & 1.2        & 1,600              & 7,700       & 500           &  27         \\ 
300 & 2  & 5 & 0       & 3.4    & 15         & 19,000             & 26,000      & 19,000        &  3,300       \\ 
50  & 2  & 3.5 & 0     & 2.5    & 12         & 25,000             & 18,000      & 9,300         &  4,100       \\ 
50$^*$ & 2 & 3.5 & 0   & 2.5    & 11         & 13,000             & 17,000      & 8,500         &  3,300       \\
200    &3.5& 5.0 & 0   & 1.9    & 2.0        & 2,400              & 12,000      & 2,900         &  230         \\
200$^*$ &3.5& 5.0 & 0  & 2.1    & 3.0        & 3,400              & 11,000      & 2,900         &  220         \\
 
\hline
\end{tabular}
\end{center}
\caption{A summary of different models. See main text for the definition of parameters shown here. 
The asterisks denote the cases, where only particles between $D=10$ $\mu$m and $D=3$~mm were considered.}
\end{table}

\clearpage

\clearpage

\clearpage

\clearpage

\clearpage

\clearpage

\clearpage

\clearpage

\clearpage

\clearpage

\clearpage

\clearpage

\clearpage

\clearpage

\clearpage

\clearpage

\clearpage

\clearpage

\end{document}